\documentclass[11pt,twoside]{article}
\usepackage{times,amssymb,cite,graphics}
\usepackage{amsthm}
\newcommand{\be}{\begin{eqnarray}}
\newcommand{\ee}{\end{eqnarray}}
\newcommand{\bez}{\begin{eqnarray*}}
\newcommand{\eez}{\end{eqnarray*}}
\newcommand{\pa}{\partial}
\newcommand{\cA}{\mathcal{A}}
\renewcommand{\d}{\mathrm{d}}
\newcommand{\tr}{\mathrm{tr}}
\newcommand{\tpsi}{\tilde{\psi}}
\newcommand{\tw}{\tilde{w}}
\newcommand{\txi}{\tilde{\xi}}
\newcommand{\res}{\mathrm{res}}
\newcommand{\bt}{\mathbf{t}}
\newcommand{\tbt}{\mathbf{t}_o}
\newcommand{\tbs}{\mathbf{s}_o}

\newtheoremstyle{noitalics}{\topsep}{\topsep}%
     {}
     {}
     {\bfseries}
     {}
     { }
     {\thmname{#1}\thmnumber{ #2}\thmnote{ #3}}    
\theoremstyle{noitalics}

\title{\bf BKP and CKP revisited: 
The odd KP system\thanks{\copyright 2008 by A. Dimakis and F. M\"uller-Hoissen}}

\author{Aristophanes Dimakis \\
 Department of Financial and Management Engineering, \\
 University of the Aegean, 31 Fostini Str., GR-82100 Chios, Greece \\
 dimakis@aegean.gr
          \and
 Folkert M\"uller-Hoissen \\
 Max-Planck-Institute for Dynamics and Self-Organization \\
 Bunsenstrasse 10, D-37073 G\"ottingen, Germany \\
 folkert.mueller-hoissen@ds.mpg.de }

\date{ }

\begin{document}

\renewcommand{\theequation} {\arabic{section}.\arabic{equation}}

\maketitle

\newtheorem{theorem}{Theorem}[section]
\newtheorem{lemma}{Lemma}[section]
\newtheorem{proposition}{Proposition}[section]
\newtheorem{example}{Example}[section]
\newtheorem{remark}{Remark}[section]
\newtheorem{corollary}{Corollary}[section]

\begin{abstract}
Restricting a linear system for the KP hierarchy to those independent variables $t_n$ 
with odd $n$, its compatibility (Zakharov-Shabat conditions) leads to the 
``odd KP hierarchy''. The latter consists of pairs of equations for \emph{two} dependent 
variables, taking values in a (typically noncommutative) associative algebra. 
If the algebra is commutative, the odd KP hierarchy is known to admit reductions to 
the BKP and the CKP hierarchy. 
We approach the odd KP hierarchy and its relation to BKP and CKP in different ways, 
and address the question whether noncommutative versions of the BKP and the CKP equation 
(and some of their reductions) exist.
In particular, we derive a functional representation of a linear system for the 
odd KP hierarchy, which in the commutative case produces functional representations 
of the BKP and CKP hierarchies in terms of a tau function. 
Furthermore, we consider a functional representation of the KP hierarchy that involves a second 
(auxiliary) dependent variable and features the odd KP hierarchy directly as a subhierarchy. 
A method to generate large classes of exact solutions to the KP hierarchy from solutions 
to a linear matrix ODE system, via a hierarchy of matrix Riccati equations, then also 
applies to the odd KP hierarchy, and this in turn can be exploited, in particular, 
to obtain solutions to the BKP and CKP hierarchies. 
\end{abstract}

\section{Introduction}
\setcounter{equation}{0}
Many (e.g. in the sense of the inverse scattering method) ``integrable'' 
 partial differential (or difference) equations (PDEs) admit generalizations 
to versions where the dependent variable takes values in an arbitrary 
associative and typically \emph{noncommutative} algebra (provided that 
 differentiability with respect to the independent variables can be 
defined) (see e.g. \cite{March88,Dorf+Foka92,Olve+Soko98assoc,Kupe00}). This fact can be 
exploited to generate large classes of exact solutions 
to a scalar integrable PDE via simple solutions to 
the corresponding matrix PDE (see also \cite{Carl+Schi99,DMH08bidiff}). 
In particular, the existence of families of solutions like multi-solitons is 
then a consequence of the existence of certain solutions to the matrix PDE 
universally for arbitrary matrix size.

There are, however, integrable equations that do \emph{not} admit a direct 
noncommutative generalization in the above sense. 
The Sawada-Kotera equation \cite{Sawa+Kote74} belongs to these exceptions 
\cite{Olve+Soko98assoc}. This equation is a reduction of the BKP equation, 
the first member of the BKP hierarchy 
\cite{DKM81TII,DJKM82TIV,DJKM81TVI,DJKM82RIMS,DJKM82TV,DJM83,DKJM83} (see also 
\cite{Kono+Dubr84,Hiro86,Hiro89a,Hiro89b,Shiota89,Nimm90,Gils+Nimm90,Hiro+Ohta91,You91,Taim91,Chow+Dasg91a,Chow+Dasg91b,Jarv+Yung93,BJY93,Naka+Yama94,vandL95,Nimm95,Plant+Salam95,Tsuj+Hiro96,Kac+vanderLeur98,vandL01,Mana+Mart98,Plaz98,Liu+Mana98,DMM99,Lori+Willo99,Lori01,Gils02,DFL02,Dubr+Lisi02,Muse+Verh03,Orlov03,Hiro04,Nimm+Orlo05,Kric06,HCR06,Hu+Wang06,Taka06sig,Taka07,Tu07,HWC07,vandL+Orlov08}),
which also lacks a noncommutative version (the latter should not be confused 
with the multi-component version of BKP). 
The BKP hierarchy and also the CKP hierarchy \cite{DJKM81TVI,DJKM82TIV} 
(see also \cite{Kono+Dubr84,Nimm95,Lori99,Lori01,DFL02,Dubr+Lisi02,Muse+Verh03,Arat+vandeL05,Matsuno05DP1,HCR06,HCR07,HWC07}) 
originate from the ``commutative'' KP hierarchy in the Gelfand-Dickey-Sato (GDS) formalism 
(see section~\ref{subsec:GDS}) by first restricting the Lax equations to only odd-numbered variables
$t_1,t_3,t_5, \ldots$, and then imposing additional reduction conditions. 
The first step clearly also works in the noncommutative case. It leads to 
the (noncommutative) ``odd KP hierarchy''. 

The GDS formulation of the KP hierarchy involves an infinite number of dependent 
variables. All besides one can be eliminated, resulting in PDEs for a single dependent variable. 
In the same way, the odd KP hierarchy (in the GDS formalism) leaves us with 
PDEs for \emph{two} dependent variables. These PDEs admit symmetries by means of which the full 
KP hierarchy can be restored (and the two dependent variables reduced to a single one). 
This shows that the odd KP hierarchy is a part (subhierarchy) of 
the KP hierarchy, something that is obvious in its GDS form. So why should we deal with a 
subhierarchy if we could treat the full hierarchy? The crucial point is that the BKP and CKP 
reductions of the odd KP hierarchy are \emph{not} compatible with the abovementioned KP-restoring 
symmetries. The general message is that a subhierarchy can admit a reduction that does not extend 
to a reduction of the full hierarchy. And this is the reason why BKP and CKP retain their 
individuality, despite their KP origin. 

In section~\ref{sec:oddKP} we derive the first member of the odd KP hierarchy in an elementary way. 
This ``odd KP system'' is a system of two 
PDEs for the KP variable and one additional dependent variable.\footnote{Throughout we will work with 
a potential $\phi$ related to the KP variable $u$ by $u = \phi_{t_1}$, hence this system may rather 
be called ``potential odd KP system''. }
Within this system we can then look for noncommutative versions of reductions known in the 
commutative case, and this is done in some subsections of section~\ref{sec:oddKP}. 
BKP and CKP possess a certain noncommutative extension with a single dependent variable, but severely constrained. 
It turns out, in particular, that these extensions are solved 
by any solution to the first two equations of the ``noncommutative'' (potential) KdV hierarchy, 
and this result remains true in the commutative case (where the constraints disappear). 
Furthermore, there is a natural noncommutative generalization of the CKP equation, though as a system with two 
dependent variables. Nothing similar is found in the BKP case. 

In section~\ref{sec:func_linsys} we derive a linear system, in functional form, for 
the whole odd KP hierarchy and deduce corresponding results for the BKP and CKP hierarchies. 
Section~\ref{sec:funct} takes a different route, starting from a functional representation of the 
KP hierarchy that involves an auxiliary dependent variable \cite{DMH07Burgers}. In this formulation, 
the odd KP hierarchy appears as the subhierarchy that consists of equations containing 
only partial derivatives with respect to the odd-numbered variables, $t_1,t_3,t_5,\ldots$. 
The auxiliary dependent variable then takes the role of the second dependent variable of 
the odd KP system. 
A certain symmetry reduction for the (odd) KP hierarchy is then introduced, which plays a 
crucial role in the step from odd KP to BKP and CKP.

Several classes of solutions to the matrix KP hierarchy and, if a rank one 
condition holds (see e.g. \cite{Gekh+Kasm06}), then also the scalar KP 
hierarchy, can be obtained from solutions to a system of 
\emph{linear} matrix ordinary differential equations, via a system 
of matrix Riccati equations \cite{DMH06nahier,DMH07Burgers,DMH07Ricc,DMH07Wronski}. 
This is a finite-dimensional version of the famous Sato theory for the 
KP hierarchy. Using the abovementioned formulation of the KP hierarchy that exhibits the 
odd KP hierarchy directly as a subhierarchy, this immediately also generates solutions to the odd KP hierarchy. 
This is elaborated in section~\ref{sec:oddKPsol}. Furthermore, we show how solutions to the BKP and CKP 
hierarchies can be obtained from solutions to the matrix odd KP hierarchy. 
Some final remarks are collected in section~\ref{sec:conclusions}.

\section{The odd KP system}
\label{sec:oddKP}
\setcounter{equation}{0}
The Kadomtsev-Petviashvili (KP) hierarchy (see e.g. \cite{Dick03}) is given by 
the integrability (or zero curvature) conditions 
\be
     B_{m,t_n} - B_{n,t_m} + [B_m , B_n ] = 0  \label{B_n-zerocurv}
\ee
of the linear system \cite{Kash+Miwa81}
\be
    \psi_{t_n} = B_n \, \psi \, , \qquad \quad n= 1,2,3, \ldots \, , 
             \label{KP-linsys}
\ee
where
\be
    B_n = \pa^n + \sum_{k=0}^{n-2} b_{n,k} \, \pa^k \; .  \label{B_n}
\ee
Here $\pa$ is the operator of partial differentiation with respect to 
the variable $t_1$ (hence the first of equations (\ref{KP-linsys}) is 
trivially satisfied), and $\psi_{t_n}$ denotes the partial derivative 
of $\psi$ with respect to the variable $t_n$. 
The objects $b_{n,k}$ are differentiable\footnote{This requires some additional 
structure that we need not specify here. If $\cA$ is an algebra of real or 
complex matrices, the usual differential structure will be assumed. } 
functions of $\bt=(t_1,t_2,\ldots)$ with values in some associative algebra $\cA$, 
and $\psi$ is an element of a left $\cA$-module. Correspondingly, the 
dependent variable of the ``noncommutative'' KP hierarchy is an 
$\cA$-valued function.

If $\cA$ is commutative, restricting (\ref{KP-linsys}) to only odd values of $n$, setting 
$b_{n,0}=0$ for $n=1,3,5,\ldots$, and ``freezing'' the variables $t_2,t_4,\ldots$, 
leads to the BKP hierarchy \cite{DJKM82TIV,DJKM81TVI}. 

In the following we also restrict (\ref{KP-linsys}) to only odd values of $n$, but 
do not impose further conditions right away (see also \cite{DJKM81TVI} for the 
commutative case). Section~\ref{subsec:integr-cond} derives 
the ``odd KP system'' from (\ref{KP-linsys}) with $n=3,5$ in a direct way. 
Section~\ref{subsec:GDS} identifies it as the first non-trivial member of 
the GDS formulation of the KP hierarchy, restricted to odd-numbered evolution variables. 
In sections~\ref{subsec:BKP&CKP}-\ref{subsec:red-oddKP} we consider some reductions
of the odd KP system.

\subsection{Elementary derivation of the odd KP system} 
\label{subsec:integr-cond}
Let us consider the first two non-trivial equations of the above linear system with 
odd $n$, i.e.
\be
   \psi_{t_3} &=& (\pa^3 + b_{3,1} \, \pa + b_{3,0} ) \, \psi \, , 
                        \label{psi_t3}  \\
   \psi_{t_5} &=& (\pa^5 + b_{5,3} \, \pa^3 + b_{5,2} \, \pa^2 + b_{5,1} \, \pa + b_{5,0} ) \, \psi  \; .
\ee
By exploiting the integrability condition and introducing \emph{potentials} $\phi$ and $\theta$ 
via\footnote{The shift by $\frac{3}{2} \, \phi_{t_1t_1}$ leads to a more `symmetric' form of 
the resulting equations (\ref{oddKP1}) and (\ref{oddKP2}). } 
\be
   b_{3,0} = 3 \, \theta_{t_1} + \frac{3}{2} \, \phi_{t_1t_1} \, , \qquad  
   b_{3,1} = 3 \, \phi_{t_1}     \, ,  \label{pot_phi,theta}
\ee
the coefficients of the linear system are fixed in terms of $\phi$ and $\theta$, 
\be
   \psi_{t_3} &=& ( \pa^3 + 3 \, \phi_{t_1} \pa + 3 \, \theta_{t_1} 
                  + \frac{3}{2} \, \phi_{t_1t_1} ) \, \psi \, , \label{oddKP_psi_t3}  \\
   \psi_{t_5} &=& \Big(  \pa^5 + 5 \, \phi_{t_1} \pa^3 + 5 \, ( \theta_{t_1} 
                  + \frac{3}{2} \phi_{t_1t_1} ) \, \pa^2 
                  + 5 \, ( \theta_{t_1t_1} + \frac{1}{3} \phi_{t_3} 
                  + \frac{7}{6} \phi_{t_1t_1t_1}   \nonumber \\
              && + \phi_{t_1}{}^2 )  \, \pa + b_{5,0} \Big) \, \psi \, ,  \label{oddKP_psi_t5}
\ee
where
\be
 b_{5,0} &=& \frac{5}{3} \, \theta_{t_3} + \frac{10}{3} \, \theta_{t_1t_1t_1} 
         + \frac{5}{6} \, \phi_{t_1t_3} + \frac{5}{3} \, \phi_{t_1t_1t_1t_1}
         + 5 \, \{ \theta_{t_1} \, , \, \phi_{t_1} \} 
         + \frac{5}{2} \, (\phi_{t_1}{}^2)_{t_1} 
                                   \nonumber \\
        & &  + \frac{5}{3} \, [ \phi_{t_1} \, , \, \phi_{t_1t_1} ]
             + \frac{5}{3} \int [ \phi_{t_3} \, , \, \phi_{t_1} ] \, \d t_1 \; .
\ee
Here $[ \, , \, ]$ and $\{ \, , \, \}$ mean commutator and anti-commutator, respectively.
The remaining integrability conditions then result in the following pair of equations,
\be
 && \Big( 9 \, \phi_{t_5}  - 5 \, \phi_{t_1t_1t_3} + \phi_{t_1t_1t_1t_1t_1} 
    - \frac{15}{2} \{ \phi_{t_1} \, , \, \phi_{t_3} - \phi_{t_1t_1t_1} - \phi_{t_1}{}^2 \} 
    + \frac{45}{4} \, ( \phi_{t_1t_1}{}^2 - 4 \, \theta_{t_1}{}^2 ) \Big)_{t_1} \nonumber \\
 && - 5 \, \phi_{t_3t_3} 
    + 15 \Big( [ \phi_{t_1} , \theta_{t_3} - \theta_{t_1t_1t_1} ] 
    + [ \theta_{t_1} , \phi_{t_3} + \frac{1}{2} \phi_{t_1t_1t_1} ] 
    + \frac{3}{2} \, [ \theta_{t_1t_1} \, , \, \phi_{t_1t_1} ]  \nonumber \\
 && + \, [ \phi_{t_1} \, , \, \int [ \phi_{t_3} \, , \, \phi_{t_1} ] \, \d t_1 ] 
     \Big) = 0 \, ,     \label{oddKP1}
\ee
and 
\be
 && \Big[ 9 \, \theta_{t_5} - 5 \, \theta_{t_1t_1t_3} 
    + \theta_{t_1t_1t_1t_1t_1} 
    + \frac{15}{2} \, \Big( - \{ \phi_{t_3} \, , \, \theta_{t_1} \}
    + \{ \theta_{t_1t_1t_1} \, , \, \phi_{t_1} \} 
    + \frac{1}{2} \, \{ \theta_{t_1} \, , \, \phi_{t_1t_1} \}_{t_1} 
     \nonumber \\
 && + 6 \, \phi_{t_1} \, \theta_{t_1} \, \phi_{t_1} 
    + \frac{1}{6} \, [ \phi_{t_3} \, , \, \phi_{t_1t_1} ] 
    + \frac{1}{6} \, [ \phi_{t_1} \, , \, \phi_{t_1t_1} ]_{t_1t_1}
    - \frac{1}{4} \, [ \phi_{t_1t_1} \, , \, \phi_{t_1t_1t_1} ] \Big) \, \Big]_{t_1}
                          \nonumber \\
 && - 5 \, \theta_{t_3t_3} - \frac{15}{2} \, \{ \theta_{t_1} \, , \, \phi_{t_1} \}_{t_3} 
    + 15 \, [ \theta_{t_1} \, , \, \theta_{t_3} + \frac{1}{2} \theta_{t_1t_1t_1} 
    + \int [ \phi_{t_3} \, , \, \phi_{t_1} ] \, \d t_1 \, ]
     \nonumber \\
 && + 45 \, [ (\theta_{t_1})^2 \, , \, \phi_{t_1} ] 
    + 15 \, [ \phi_{t_1t_1} \, , \, [ \phi_{t_1} \, , \, \theta_{t_1} ] \, ]
    + \frac{15}{2} \, [ \, [ \theta_{t_1} \, , \, \phi_{t_1t_1} ] \, , \, \phi_{t_1} ]
    + \frac{25}{4} \, [ \phi_{t_1t_1t_3} \, , \, \phi_{t_1} ]
      \nonumber \\
 && - 5 \, \Big( \int [ \phi_{t_3} \, , \, \phi_{t_1} ] \, \d t_1 \Big)_{t_3}
    + \frac{15}{2} \, [ \phi_{t_3} - \phi_{t_1t_1t_1} \, , \, (\phi_{t_1})^2 ] 
    + \frac{45}{4} \, [ \phi_{t_1} \, , \, (\phi_{t_1t_1})^2 ]
    = 0    \; .    \label{oddKP2}
\ee
In the following we refer to (\ref{oddKP1}) and (\ref{oddKP2}) as the 
``odd KP system''. We note that by introducing 
\be
  \tilde{\theta} := \theta 
    + \frac{1}{2} \int [ \phi \, , \, \phi_{t_1} ] \; \d t_1 
    \, , \label{theta_transf}
\ee
which implies $\theta_{t_3} = \tilde{\theta}_{t_3} - \frac{1}{2} \, [ \phi \, , \, \phi_{t_3} ] 
- \int [ \phi_{t_3} \, , \, \phi_{t_1} ] \; \d t_1$, the resulting equations 
no longer involve integrals, see also section~\ref{sec:funct}.

\begin{remark}
\label{rem:oddKP->KP}
Switching on ``even flows'', we have in particular 
$\psi_{t_2} = (\pa^2 + b_{2,0}) \psi$. Compatibility with 
(\ref{psi_t3}) (using (\ref{pot_phi,theta})) then leads to
$b_{2,0} = 2 \, \phi_{t_1}$, $\theta_{t_1} = \frac{1}{2} \phi_{t_2}$, 
and the (potential) KP equation for $\phi$. 
\hfill $\square$ 
\end{remark}

\subsection{Recovering BKP and CKP in the commutative case} 
\label{subsec:BKP&CKP}
If $\cA$ is commutative, then the above pair of equations reduces to
\be
 && \Big( 9 \, \phi_{t_5}  - 5 \, \phi_{t_1t_1t_3} + \phi_{t_1t_1t_1t_1t_1} 
    - 15 \, \phi_{t_3} \, \phi_{t_1}  + 15 \, \phi_{t_1} \, \phi_{t_1t_1t_1} 
    + 15 \, \phi_{t_1}{}^3 
    + \frac{45}{4} \, \phi_{t_1t_1}{}^2  
            \nonumber \\
 && - 45 \, \theta_{t_1}{}^2 \Big)_{t_1} 
    - 5 \, \phi_{t_3t_3} = 0 \, ,     \label{oddKP1c}  \\
 && \Big( 9 \, \theta_{t_5}  - 5 \, \theta_{t_1t_1t_3} 
    + \theta_{t_1t_1t_1t_1t_1} 
    - 15 \, \phi_{t_3} \, \theta_{t_1}  
    + 15 \, \phi_{t_1} \, \theta_{t_1t_1t_1}  
    + \frac{15}{2} \, ( \phi_{t_1t_1} \, \theta_{t_1} )_{t_1} 
     \nonumber \\
 && + 45 \, \phi_{t_1}{}^2 \, \theta_{t_1}   \Big)_{t_1}
    - 5 \, ( \theta_{t_3} + 3 \, \phi_{t_1} \, \theta_{t_1} )_{t_3}
    = 0    \; .    \label{oddKP2c}
\ee
Setting
\be
     \theta = k \, \phi_{t_1} \, ,   \label{theta=kphi_t1}
\ee
it turns out that the second equation is a consequence of the first if 
\be
      k = 0, \pm \frac{1}{2} \; .
\ee

If $k= \pm 1/2$, (\ref{oddKP1c}) becomes the \emph{BKP equation} 
\be
  \Big( 9 \, \phi_{t_5}  - 5 \, \phi_{t_1t_1t_3} + \phi_{t_1t_1t_1t_1t_1} 
    - 15 \, \phi_{t_1} \, \phi_{t_3} + 15 \, \phi_{t_1} \, \phi_{t_1t_1t_1}  
    + 15 \, \phi_{t_1}{}^3 \Big)_{t_1}  - 5 \, \phi_{t_3t_3}
     = 0 \; .  \qquad   \label{BKP} 
\ee
Setting $\phi_{t_3} = 0$ reduces (\ref{BKP}) to the (potential) Sawada-Kotera 
equation \cite{Sawa+Kote74,DJKM82TIV,Hiro04}
\be
      9 \, \phi_{t_5} + \phi_{t_1t_1t_1t_1t_1} 
    + 15 \, ( \phi_{t_1} \, \phi_{t_1t_1t_1} + \phi_{t_1}{}^3 )
     = 0 \, ,     \label{Sawada-Kotera} 
\ee
which is known \emph{not} to possess a \emph{non}commutative (e.g. matrix) version \cite{Olve+Soko98assoc}. 
Setting $\phi_{t_5} = 0$ in (\ref{BKP}), yields the Ramani equation \cite{Rama81,DJKM82TIV} 
(also called (potential) bidirectional Sawada-Kotera (bSK) equation \cite{Dye+Park01,Dye+Park02,Park+Dye02,HCR06}),
\be
   \Big( - 5 \, \phi_{t_1t_1t_3} + \phi_{t_1t_1t_1t_1t_1} 
    - 15 \, \phi_{t_1} \, \phi_{t_3} + 15 \, \phi_{t_1} \, \phi_{t_1t_1t_1} 
    + 15 \,  \phi_{t_1}{}^3 \Big)_{t_1}
    - 5 \, \phi_{t_3t_3}
     = 0 \; .  \qquad   \label{Ramani} 
\ee
\vskip.1cm

If $k=-1/2$ we have $b_{3,0}=b_{5,0}=0$ and thus the familiar linear system for the 
BKP equation \cite{DJKM82TIV,DJKM81TVI}, 
\be
  \psi_{t_3} &=& (\pa^3 + 3 \, \phi_{t_1} \, \pa) \, \psi \, , \\
  \psi_{t_5} &=& \Big( \pa^5  + 5 \, \pa \, \phi_{t_1} \, \pa^2 
     + \frac{5}{3} \, ( \phi_{t_3} + 2 \, \phi_{t_1t_1t_1} + 3 \, \phi_{t_1}{}^2 ) 
     \, \pa \Big) \, \psi \; .
\ee
\vskip.1cm

If $k=1/2$, we obtain another linear system for the BKP equation:
\be
  \psi_{t_3} &=& 3 \, \phi_{t_1t_1} \, \psi + 3 \, \phi_{t_1} \, \psi_{t_1} 
                   + \psi_{t_1t_1t_1} 
              = ( \pa^3 + 3 \, \pa \, \phi_{t_1} ) \, \psi      \, , \\
  \psi_{t_5} &=& \frac{5}{3} \, ( \phi_{t_3} + 2 \, \phi_{t_1t_1t_1} 
                 + 3 \, \phi_{t_1}{}^2 )_{t_1} \, \psi  
                 + \frac{5}{3} \, ( \phi_{t_3} + 5 \, \phi_{t_1t_1t_1} 
                 + 3 \, \phi_{t_1}{}^2 ) 
                       \, \psi_{t_1}  \nonumber \\
             & & + 10 \, \phi_{t_1t_1} \, \psi_{t_1t_1} 
                 + 5 \, \phi_{t_1} \, \psi_{t_1t_1t_1} + \psi_{t_1t_1t_1t_1t_1} 
                   \nonumber \\
             &=& \Big( \pa^5 + 5 \, \pa^2 \phi_{t_1} \pa + \frac{5}{3} \, \pa \, ( \phi_{t_3} 
                 + 2 \, \phi_{t_1t_1t_1} + 3 \, \phi_{t_1}{}^2 ) \Big) \, \psi   \, ,
\ee
which is thus simply an adjoint of the first linear system. 
\vskip.1cm

If $k=0$ (i.e. $\theta=0$), (\ref{oddKP1c}) becomes the \emph{CKP equation} \cite{DJKM82TIV}  
\be
 && \Big(  9 \, \phi_{t_5}  - 5 \, \phi_{t_1t_1t_3} + \phi_{t_1t_1t_1t_1t_1} 
    - 15 \, \phi_{t_1} \, \phi_{t_3} + 15 \, \phi_{t_1} \, \phi_{t_1t_1t_1} 
            + 15 \, \phi_{t_1}{}^3  
    + \frac{45}{4} \, \phi_{t_1t_1}{}^2 \Big)_{t_1} \nonumber \\
 && - 5 \, \phi_{t_3t_3}
     = 0 \; .  \qquad   \label{CKP} 
\ee
The linear system in this case turns out to be given by half the sum of the respective 
equations of the above two BKP linear systems. 

Setting $\phi_{t_3} = 0$ reduces (\ref{CKP}) to the (potential) Kaup-Kupershmidt 
equation \cite{Kaup80}
\be
   9 \, \phi_{t_5} + \phi_{t_1t_1t_1t_1t_1} 
    + 15 \, ( \phi_{t_1} \, \phi_{t_1t_1t_1}   
    + \frac{3}{4} \, \phi_{t_1t_1}{}^2 + \phi_{t_1}{}^3 )
     = 0 \; .  \qquad   \label{KK} 
\ee
Setting $\phi_{t_5} = 0$ in (\ref{CKP}), yields the bidirectional Kaup-Kupershmidt (bKK) equation
\cite{Dye+Park01,Dye+Park02,Park+Dye02,Verh+Muse03,HCR06}.

\subsection{BKP and the noncommutative KdV hierarchy} 
\label{subsec:BKP&ncKdV}
Imposing the BKP condition  
\be
    \theta = - \frac{1}{2} \, \phi_{t_1}  \label{theta=-phit1/2}
\ee
(i.e. $k=- \frac{1}{2}$ in (\ref{theta=kphi_t1})) in the \emph{non}commutative case, 
we have $b_{3,0} =0$ and 
\be
  b_{5,0} = \frac{5}{3} \int [ \phi_{t_3} - \phi_{t_1t_1t_1} \, , \, \phi_{t_1} ]
            \, \d t_1 \; .  \label{b50_ncBKP}
\ee
Then (\ref{oddKP1}) reduces to  
\be
 && \Big( 9 \, \phi_{t_5}  - 5 \, \phi_{t_1t_1t_3} 
    + \phi_{t_1t_1t_1t_1t_1} 
    - 15 \, ( \phi_{t_1} \, ( \phi_{t_3} - \phi_{t_1t_1t_1} ) 
    - \phi_{t_1}{}^3 ) \Big)_{t_1}  - 5 \, \phi_{t_3t_3}           \nonumber \\
 && + 15 \, [ \phi_{t_1} \, , \, \int [ \phi_{t_3} \, , \, \phi_{t_1} ] 
    \, \d t_1 ] 
    = 0 \, ,     \label{ncBKP}
\ee
and (\ref{oddKP2}), after use of (\ref{ncBKP}), becomes
\be
 [ \phi_{t_3} - \phi_{t_1t_1t_1} \, , \, \phi_{t_1} ]_{t_1t_1} 
    - \Big( \int [ \phi_{t_3} - \phi_{t_1t_1t_1} \, , \, \phi_{t_1} ] 
      \, \d t_1 \Big)_{t_3}
    + 3 \, \phi_{t_1} \, [ \phi_{t_3} - \phi_{t_1t_1t_1} \, , \, \phi_{t_1} ]  
  = 0 \; .         \label{ncBKP_constr}
\ee
The latter equation represents a constraint which, however, is \emph{not} in general 
preserved under the flow with evolution variable $t_5$, given by (\ref{ncBKP}).\footnote{Taking 
a look at this problem in the Sawada-Kotera case, where $\phi_{t_3} =0$ simplifies the equations 
a lot, we have to compute the derivative of (\ref{ncBKP_constr}) with respect to $t_5$ and then 
eliminate $\phi_{t_5}$ by use of (\ref{ncBKP}). Already the resulting terms quadratic in 
(derivatives of) $\phi$ do not cancel as a consequence of (\ref{ncBKP_constr}) and its derivatives 
with respect to $t_1$. }
Now we observe that (\ref{ncBKP_constr}) is obviously solved if 
\be
   \phi_{t_3} = \phi_{t_1t_1t_1} + f(\phi_{t_1}) \, ,   \label{gKdV}
\ee
where $f$ is an arbitrary polynomial in $\phi_{t_1}$ with coefficients in the center 
of $\cA$. But only for a special choice of $f$, the equation (\ref{gKdV}) 
has a chance to be compatible with (\ref{ncBKP}). 
Addressing integrability, evolution equations like (\ref{gKdV}), with the restriction 
that the right hand side is a homogeneous differential polynomial, appear to be 
distinguished. This reduces (\ref{gKdV}) to the potential KdV equation, where  
$f(\phi_{t_1}) = a \, \phi_{t_1}{}^2$ with a constant $a$, or the mKdV equation, 
where $f(\phi_{t_1}) = a \, \phi_{t_1}{}^3$. But only in the KdV case the 
weighting of terms is compatible with (\ref{ncBKP}). 
Using the KdV equation in (\ref{ncBKP}), yields
\be
  && \Big( 9 \, \phi_{t_5} - 9 \, \phi_{t_1t_1t_1t_1t_1} 
   - 15 \, a \, ( \phi_{t_1}{}^2 )_{t_1t_1} + 15 \, a \, \phi_{t_1t_1}{}^2  
   - 5 \, (a+3) \, a \, \phi_{t_1}{}^3 \Big)_{t_1}  \nonumber \\
  &&  + 5 \, (9 - a^2) \, \phi_{t_1} \phi_{t_1t_1} \phi_{t_1} = 0 \; .
\ee
Choosing $a=3$, this can be integrated to
\be
   \phi_{t_5} - \phi_{t_1t_1t_1t_1t_1} - 5 \, (\phi_{t_1}{}^2)_{t_1t_1} 
   + 5 \, \phi_{t_1t_1}{}^2 - 10 \, \phi_{t_1}{}^3 = 0 \, ,  \label{ncpKdV2_BKP}
\ee
and (\ref{gKdV}) reads
\be
    \phi_{t_3}  = \phi_{t_1t_1t_1} + 3 \, \phi_{t_1}{}^2 \; .   \label{ncpKdV1_BKP}
\ee
(\ref{ncpKdV1_BKP}) and (\ref{ncpKdV2_BKP}) are the first two equations of the 
noncommutative potential KdV (ncpKdV) hierarchy.\footnote{With $u = - \phi_x$ 
where $x=t_1$ we obtain from (\ref{ncpKdV1_BKP}) and (\ref{ncpKdV2_BKP}), respectively, 
the potential versions of (3.46) and (3.47) in \cite{DMH04hier}.}
Hence, any solution to the first two ncpKdV hierarchy 
equations (\ref{ncpKdV1_BKP}) and (\ref{ncpKdV2_BKP}) also solves the above 
noncommutative extension of the BKP equation.\footnote{An analogous relation 
exists between the first two equations of the (noncommutative) Burgers hierarchy and 
the KP equation \cite{MSS91,DMH07Burgers}.}
This relation then also holds for the ``commutative'' scalar equations, of course.  
But to find this result the step into the noncommutative framework was extremely helpful.

\begin{remark}
If we impose the conditions $b_{3,0} = b_{5,0} = 0$ on the noncommutative odd KP system, 
we obtain (\ref{theta=-phit1/2}) and 
$[ \phi_{t_3} - \phi_{t_1t_1t_1} \, , \, \phi_{t_1} ] =0$, which leads more directly to (\ref{gKdV}). 
\hfill $\square$
\end{remark}
\vskip.2cm

Another noncommutative extension of the BKP equation is obtained for 
$\theta = \frac{1}{2} \, \phi_{t_1}$ (i.e. $k= \frac{1}{2}$ in (\ref{theta=kphi_t1})), 
and one finds corresponding results.

\subsection{CKP and the noncommutative KdV hierarchy} 
\label{subsec:CKP&ncKdV}
Imposing the CKP condition $\theta =0$, (\ref{oddKP1}) reduces to
\be
 && \Big( 9 \, \phi_{t_5}  - 5 \, \phi_{t_1t_1t_3} + \phi_{t_1t_1t_1t_1t_1} 
    - \frac{15}{2} \{ \phi_{t_1} \, , \, \phi_{t_3} - \phi_{t_1t_1t_1} - \phi_{t_1}{}^2 \} 
    + \frac{45}{4} \, (\phi_{t_1t_1})^2 \Big)_{t_1}
     \nonumber \\
 && - 5 \, \phi_{t_3t_3}
    + 15 \, [ \phi_{t_1} \, , \, \int [ \phi_{t_3} \, , \, \phi_{t_1} ] \, \d t_1 ] 
     = 0     \label{ncCKP1}
\ee
and (\ref{oddKP2}) yields a constraint, involving only commutators, which is not in general 
preserved under the flow of (\ref{ncCKP1}). 
The constraint turns out to be satisfied as a consequence of the ncpKdV equation in the form
\be
    \phi_{t_3} = \frac{1}{4} \, \phi_{t_1t_1t_1} + \frac{3}{2} \, (\phi_{t_1})^2 \, ,
          \label{ncpKdV1_CKP}
\ee
and (\ref{ncCKP1}) then integrates to 
\be
   \phi_{t_5} = \frac{1}{16} \, \phi_{t_1t_1t_1t_1t_1} 
      + \frac{5}{8} \, (\phi_{t_1}{}^2)_{t_1t_1} 
      - \frac{5}{8} \, (\phi_{t_1t_1})^2 
      + \frac{5}{2} \, (\phi_{t_1})^3 \, ,     \label{ncpKdV2_CKP}
\ee
which is the second equation of the ncpKdV hierarchy.\footnote{We note that 
(\ref{ncpKdV1_CKP}) and (\ref{ncpKdV2_CKP}) can be obtained from (\ref{ncpKdV1_BKP}) 
and (\ref{ncpKdV2_BKP}) via $t_n \mapsto 2 \, t_n$. }
As a consequence, any solution to the first two equations of the ncpKdV hierarchy (with 
coefficients as given above) is also a solution to the constrained noncommutative extension 
of the CKP equation. In the commutative case, the corresponding statement then also holds, 
of course, i.e. any solution to the first two equations of the potential KdV hierarchy (with 
coefficients as given above) is also a solution to the CKP equation.

\subsection{Further reductions of the odd KP system in the noncommutative case}
\label{subsec:red-oddKP}
Imposing $\phi_{t_3} = \theta_{t_3} =0$, we obtain from (\ref{oddKP1}) 
and (\ref{oddKP2}) the following noncommutative generalization of the 
(potential) Sawada-Kotera (\ref{Sawada-Kotera}) and Kaup-Kupershmidt 
equation (\ref{KK}),
\be
 &&  9 \, \phi_{t_5} + \phi_{t_1t_1t_1t_1t_1} 
    + \frac{15}{2} \, \{ \phi_{t_1} \, , \, \phi_{t_1t_1t_1} \}  
    + \frac{45}{4} \, \phi_{t_1t_1}{}^2 + 15 \, \phi_{t_1}{}^3
    + 15 \, [ \theta_{t_1t_1} \, , \, \phi_{t_1} ]           \nonumber \\
 &&  + \frac{15}{2} \, [ \theta_{t_1} \, , \, \phi_{t_1t_1} ] - 45 \, \theta_{t_1}{}^2 
     = 0        \label{SK1}
\ee
and 
\be
 &&  9 \, \theta_{t_1t_5} + \theta_{t_1t_1t_1t_1t_1t_1} 
    + \frac{15}{2} \Big( \{ \theta_{t_1t_1t_1} \, , \, \phi_{t_1} \} 
      + \frac{1}{2} \, \{ \theta_{t_1} \, , \, \phi_{t_1t_1} \}_{t_1} 
      + 6 \, \phi_{t_1} \, \theta_{t_1} \, \phi_{t_1} 
      + \frac{1}{6} \, [ \phi_{t_1} \, , \, \phi_{t_1t_1} ]_{t_1t_1}
                          \nonumber \\
 && - \frac{1}{4} \, [ \phi_{t_1t_1} \, , \, \phi_{t_1t_1t_1} ] 
  + [ \theta_{t_1} \, , \,  \theta_{t_1t_1t_1} ] \Big)_{t_1} 
  + 45 \, [ (\theta_{t_1})^2 - \frac{1}{4} (\phi_{t_1t_1})^2 \, , \, \phi_{t_1} ] 
  + 15 \, [ \phi_{t_1t_1} \, , \, [ \phi_{t_1} \, , \, \theta_{t_1} ] \, ]
     \nonumber \\
 && + \frac{15}{2} \, [ \, \phi_{t_1}\, , \, [ \phi_{t_1t_1} \, , \, \theta_{t_1} ] \,]
    - \frac{15}{2} \, [ \phi_{t_1t_1t_1} \, , \, (\phi_{t_1})^2 ] 
    = 0    \; .    \label{SK2}
\ee
In the \emph{commutative} case, the last equation can be integrated with 
respect to $t_1$, and we recover an integrable system that 
appeared in \cite{Four+More01,Four+More02} (see also \cite{MSY87}), 
\be
 && 9 \, u_{t_5} + u_{t_1t_1t_1t_1t_1} 
    + 10 \, u \, u_{t_1t_1t_1}
    + 25 \, u_{t_1} \, u_{t_1t_1} + 20 \, u^2 \, u_{t_1} 
    - 135 \, \theta_{t_1} \, \theta_{t_1t_1} = 0 \, , \\
 && 9 \, \theta_{t_5} + \theta_{t_1t_1t_1t_1t_1} 
    + 10 \, u \, \theta_{t_1t_1t_1} 
    + 5 \, ( u_{t_1} \, \theta_{t_1} )_{t_1}  
    + 20 \, u^2 \, \theta_{t_1} = 0 \, , 
\ee
where $u := \frac{3}{2} \phi_{t_1}$. 
 In \cite{Four99} an attempt was made 
to find a noncommutative version of ``coupled systems of Kaup-Kupershmidt 
and Sawada-Kotera type'', but without success. The above equations 
(\ref{SK1}) and (\ref{SK2}) constitute a solution to this problem. 
\vskip.1cm

Setting $\phi_{t_5} = \theta_{t_5} =0$ in (\ref{oddKP1}) 
and (\ref{oddKP2}), we obtain a system that may be regarded as a noncommutative 
generalization of the Ramani (or bSK) equation (\ref{Ramani}) and 
the bidirectional Kaup-Kupershmidt (bKK) equation.

\begin{remark}
The system (\ref{SK1}) and (\ref{SK2}) possesses the symmetry $\phi_{t_2} = 2 \, \theta_{t_1}$
(see also remark~\ref{rem:oddKP->KP}), 
by use of which we obtain from it the first and the third member of the (noncommutative)  
Boussinesq hierarchy. The latter is the \emph{3-reduction} of the (noncommutative) KP hierarchy 
(also called third Gelfand-Dickey hierarchy \cite{Dick03}). This means that the system (\ref{SK1}) and 
(\ref{SK2}) can also be obtained as a reduction of the KP hierarchy, and not just as a 
reduction of the odd KP hierarchy. The crucial point is that the reduction condition is 
compatible with the equations (like $\phi_{t_2} = 2 \, \theta_{t_1}$) that are needed to 
complete the odd KP hierarchy to the KP hierarchy (cf section~\ref{sec:funct}). This is not so for 
the reductions of odd KP to BKP or CKP. In the same way, the noncommutative generalization of the 
bSK and bKK equations is related to the 5-reduction of the (noncommutative) KP hierarchy 
(fifth Gelfand-Dickey hierarchy).
\hfill $\square$
\end{remark}

\subsection{Gelfand-Dickey-Sato formulation of the odd KP hierarchy}
\label{subsec:GDS}
The odd KP system can be extended to a hierarchy by restricting the GDS 
formulation (see e.g. \cite{Dick03}) of the KP hierarchy, 
\be
     L_{t_n} = [ B_n , L ] \, , \label{KP-GD_Lax}
\ee
where
\be
    B_n = (L^n)_{\geq 0} \, ,  \qquad 
    L = \pa + u_2 \, \pa^{-1} + u_3 \, \pa^{-2} + \ldots \, , 
    \label{KP-GD-B_nL}
\ee
to odd-numbered variables $t_n$. Here 
$\pa^{-1}$ is the formal inverse of $\pa$ and $( \; )_{\geq 0}$ 
means the projection of a pseudodifferential operator to its 
differential operator part (see e.g. \cite{Dick03}). 
We have in particular
\be
    B_3 &=& (L^3)_{\geq 0}
         = \pa^3 + 3 \, u_2 \, \pa + 3 \, (u_3 + u_{2,t_1}) \, , \\
    B_5 &=& (L^5)_{\geq 0}
         = \pa^5 + 5 \, u_2 \, \pa^3 + 5 \, (u_3 + 2 u_{2,t_1}) \, \pa^2
            + 5 \, ( u_4 + 2 \, u_{3,t_1} + 2 \, u_{2,t_1t_1}
                     + 2 \, u_2^2) \, \pa \nonumber \\
        &&  + 5 \, ( u_5 + 2 \, u_{4,t_1} + 2 \, u_{3,t_1t_1} 
                    + u_{2,t_1t_1t_1} + 2 \, \{ u_2 , u_3 \}
                    + 2 \, (u_2^2)_{t_1} ) \; .
\ee
(\ref{KP-GD_Lax}) is known to be equivalent to the zero curvature conditions 
(\ref{B_n-zerocurv}), with $B_n$ defined in (\ref{KP-GD-B_nL}). 
By comparison with $B_3$ and $B_5$ computed in section~\ref{subsec:integr-cond}, 
we find
\be
      u_2 &\!=\!& \phi_{t_1} \, , \qquad 
      u_3 = \theta_{t_1} - \frac{1}{2} \, \phi_{t_1t_1} \, , \qquad 
      u_4 = - \theta_{t_1t_1} + \frac{1}{3} \, \phi_{t_3} 
            + \frac{1}{6} \, \phi_{t_1t_1t_1} - (\phi_{t_1})^2 \, , \nonumber \\
      u_5 &\!=\!& \frac{1}{3} \, \theta_{t_3} + \frac{2}{3} \, \theta_{t_1t_1t_1}
            - \frac{1}{2} \, \phi_{t_1t_3} - \{ \theta_{t_1} , \phi_{t_1} \}
            + \frac{3}{2} \, (\phi_{t_1}{}^2)_{t_1} 
            + \frac{1}{3} \, [ \phi_{t_1} , \phi_{t_1t_1} ]  \nonumber \\
       & &  + \frac{1}{3} \, \int [ \phi_{t_3} , \phi_{t_1} ] \, \d t_1
            \, . 
\ee 

If $\cA$ is \emph{commutative}, the CKP reduction of the KP hierarchy is 
determined by $L + L^\ast = 0$, and the BKP reductions by 
$\pa \, L + L^\ast \, \pa = 0$, respectively $L \, \pa + \pa \, L^\ast =0$
\cite{DJKM81TVI}.
Here $L^\ast$ denotes the adjoint of the pseudodifferential operator $L$ 
(see e.g. \cite{Dick03}). We summarize these well-known relations together with those 
found in section~\ref{subsec:BKP&CKP} in the following table.
\begin{center}
\begin{tabular}{|l|l|l|} \hline
  BKP & $\pa \, L + L^\ast \, \pa = 0$ & $\theta = -\frac{1}{2} \, \phi_{t_1}$  \\
  \hline 
  BKP & $ L \, \pa + \pa \, L^\ast = 0$ & $\theta = \frac{1}{2} \, \phi_{t_1}$  \\
  \hline
  CKP & $ L + L^\ast = 0$ & $\theta = 0$   \\
  \hline
\end{tabular}
\end{center}

If $\cA$ is matrix algebra over $\mathbb{R}$ or $\mathbb{C}$, we can generalize the 
adjoint by setting $(A \pa)^\ast := - \pa \, A^\intercal$, where $A \in \cA$ with transpose 
$A^\intercal$.\footnote{More generally, we may consider an algebra $\cA$ with an involution ${}^\ast$,  
and define $(A \pa)^\ast := - \pa \, A^\ast$. } 
The CKP condition then generalizes to
\begin{center}
\begin{tabular}{|l|l|l|} \hline
  matrix CKP & $ L + L^\ast = 0$ & $\phi^\intercal = \phi \, , \quad \theta^\intercal = -\theta$   \\
  \hline
\end{tabular}
\end{center}
The conditions for $\phi$ and $\theta$ indeed yield a consistent reduction of the odd KP system, 
which may thus be regarded as a noncommutative version of the CKP equation. For $m>1$, it is  
a \emph{pair} of equations for \emph{two} dependent (matrix) variables, however. The corresponding hierarchy 
will be called \emph{matrix CKP} hierarchy. In the following, ``CKP equation'' or ``CKP hierarchy'' 
throughout refers to the familiar scalar (commutative) case, i.e. $m=1$, and we will add 
``matrix'' whenever we mean the matrix generalization. 
In contrast to the CKP case, the above BKP reduction condition for $L$ does \emph{not} consistently 
generalize to the noncommutative case. 
\vskip.1cm 

The formulation (\ref{KP-GD_Lax}), with $n \in \mathbb{N}$, of the KP hierarchy depends on 
an infinite number of dependent variables. 
Elimination of $u_3,u_4,\ldots$ leads to PDEs that only involve the variable $u_2$ ($=\phi_{t_1}$).
Omitting some of the equations (\ref{KP-GD_Lax}), it will no longer be possible to 
eliminate all the auxiliary variables $u_3,u_4,\ldots$. 
In the step to the odd KP hierarchy, where all equations (\ref{KP-GD_Lax}) involving derivatives 
with respect to even-numbered variables are dropped, one of the additional variables is retained, 
namely $u_3$, which leads to the appearance of $\theta$. 
It would be desirable to find a way to explicitly eliminate all the remaining auxiliary variables 
$u_4,u_5,\ldots$ from the sequence of equations (\ref{KP-GD_Lax}) with odd $n$. In 
section~\ref{sec:func_linsys} we solve this problem on the level of the corresponding 
linear system. The odd KP hierarchy expressed in terms of $\phi$ and $\theta$ (without 
auxiliary variables) then arises from the integrability conditions. 
\vskip.1cm

Also in case of the full KP system, (\ref{KP-GD_Lax}) with $n \in \mathbb{N}$, we may think 
of eliminating only $u_4,u_5,\ldots$. The resulting equations then depend on $u_2$ and $u_3$, 
and further elimination of $u_3$ would lead to the KP equation and its companions. 
The more interesting aspect, however, is that in such a formulation of the KP hierarchy, 
we should expect the odd KP system (and its hierarchy companions) to form a subhierarchy. 
In fact, in section~\ref{sec:funct}, we start from a functional form of the KP hierarchy that 
involves one additional (auxiliary) variable that, by now not surprisingly, turns out to be 
related to $\theta$. In this representation of the KP hierarchy, the odd KP hierarchy is  
indeed nicely described as a subhierarchy. We note that in this picture a 
solution to the odd KP hierarchy in general still depends on the even-numbered variables $t_{2n}$, 
which are constants with respect to the odd KP hierarchy.

\section{A linear system for the odd KP hierarchy in functional form}
\label{sec:func_linsys}
\setcounter{equation}{0}
In this section we present a linear system for the whole noncommutative odd KP hierarchy in functional form. 
This extends the linear system for the odd KP system obtained in section~\ref{subsec:integr-cond}. 
The bilinear identity for the KP hierarchy (see e.g. \cite{Dick03}), restricted to odd-numbered variables, is
\be
       \res [\psi(\tbs,z) \, \tpsi(\tbt,z)] = 0 \, ,  \label{bilinear-id}
\ee
where $\tbt = (t_1,t_3,t_5,\ldots)$, 
\be
    \psi(\tbt,z) = w(\tbt,z) \, e^{\txi(\tbt,z)} \, , \qquad
    \tpsi(\tbt,z) = \tw(\tbt,z) \, e^{-\txi(\tbt,z)}  \, ,  \label{psi-w,tpsi-tw}
\ee
with $\txi(\tbt,z) = \sum_{n\geq 1} t_{2n-1} \, z^{2n-1}$ and
\be
    w(\tbt,z) = I + \sum_{n \geq 1} w_n(\tbt) \, z^{-n} \, ,\qquad
    \tw(\tbt,z) = I + \sum_{n \geq 1} \tw_n(\tbt) \, z^{-n} \; .
\ee
We will often omit the argument $\tbt$, for simplicity. 
Inserting (\ref{psi-w,tpsi-tw}) in (\ref{bilinear-id}), the bilinear identity reads
\be
    \res \left( w(\tbs,z) \, \tw(\tbt,z) \, e^{\txi(\tbs-\tbt,z)} \right) = 0.
    \label{bilinear-id2}
\ee
The residue $\res f(z)$ of a formal series $f(z) = \sum_{n=-\infty}^{+\infty} f_n \, z^{-n}$ is 
the coefficient $f_1$. In particular, setting $\tbs=\tbt$, (\ref{bilinear-id2}) implies
\be
    \tw_1 = -w_1 =: \phi \; .  \label{tw1=-w1}
\ee
We write
\be
    w_2 = -\tilde{\theta} + \frac{1}{2} (\phi_{t_1} + \phi^2) \, ,  \label{w2-theta}
\ee
with a variable $\tilde{\theta}$. We shall see that $\phi$ can be identified with the variable 
of the same name introduced in section \ref{subsec:integr-cond}, and that $\tilde{\theta}$ coincides 
with the variable defined in (\ref{theta_transf}). 
Below we use the Miwa shift notation $\phi_{[\lambda]}(\tbt) = \phi(\tbt + [\lambda])$, 
$[\lambda] = (\lambda, \lambda^3/3, \lambda^5/5,\ldots)$. 
The proof of the following theorem is presented in Appendix~A.

\begin{theorem}
\label{theorem:oddKP_funct_linsys}
The bilinear identity implies
\be
  &&  \frac{1}{\lambda} \, F(\lambda) \, ( \psi_{2[\lambda]} - \psi ) - (\psi_{2[\lambda]}+\psi)_{t_1}  \nonumber \\
  &&\hspace{1cm}
    = \frac{\lambda}{2} \Big(\tilde{\theta}_{2[\lambda]} - \tilde{\theta} 
      + \frac{1}{2} (\phi_{2[\lambda]} - \phi)_{t_1}
      - \frac{1}{2} \, [\phi,\phi_{2[\lambda]}] \Big) \, F(\lambda)^{-1} \, (\psi_{2[\lambda]} + \psi) \, ,
    \label{fun_oddKP}
\ee
where
\be
    F(\lambda) := I - \frac{\lambda}{2} \left( \phi_{2[\lambda]} - \phi \right) \; .
                  \label{F(la)}
\ee
\hfill $\square$
\end{theorem}

(\ref{fun_oddKP}) is a functional representation of the linear system for the odd KP hierarchy. 
By expansion in powers of the indeterminate $\lambda$, we recover from the lowest orders 
the linear system of the odd KP system derived in section~\ref{subsec:integr-cond}. 
Indeed, at order $\lambda^2$ we obtain
\be
 \psi_{t_3} = \Big( \pa^3 + 3 \, \phi_{t_1} \pa 
 + \frac{3}{2} \, ( 2 \, \tilde{\theta}_{t_1} + \phi_{t_1t_1} + [ \phi_{t_1} , \phi ] ) \Big) \, \psi \, , 
     \label{oddKP_psi_t3_tth}
\ee
which is (\ref{oddKP_psi_t3}) by use of (\ref{theta_transf}). At order $\lambda^3$ we obtain the 
derivative of the above equation with respect to $t_1$. At order $\lambda^4$ we get an equation 
that contains $\psi_{t_3}$, which can be replaced with the help of (\ref{oddKP_psi_t3_tth}). 
This results in 
\be
   \psi_{t_5} &=& \Big( \pa^5 + 5 \, \phi_{t_1} \pa^3 + \frac{5}{2} ( 2 \, \tilde{\theta}_{t_1}
     + 3 \, \phi_{t_1t_1} + [ \phi_{t_1} , \phi ] ) \, \pa^2 
     + \frac{5}{6} ( 6 \, \tilde{\theta}_{t_1t_1} + 7 \, \phi_{t_1t_1t_1} + 2 \, \phi_{t_3} 
                 \nonumber \\
    &&  + 6 \, \phi_{t_1}{}^2 + 3 \, [ \phi_{t_1t_1} , \phi ] ) \, \pa 
     + \frac{5}{3} ( \tilde{\theta}_{t_3} + 2 \, \tilde{\theta}_{t_1t_1t_1} 
                     + \phi_{t_1t_1t_1t_1} +  \frac{1}{2} \phi_{t_1t_3} )
     + 5 \, \{ \tilde{\theta}_{t_1} , \phi_{t_1} \}    \nonumber \\
    && + \frac{5}{2} \, (\phi_{t_1}{}^2)_{t_1} + \frac{5}{6} \, [ \phi_{t_3} , \phi ] 
     + \frac{5}{3} \, [ \phi_{t_1t_1t_1} , \phi ]
     + \frac{5}{2} \, [ \phi_{t_1}{}^2 , \phi ] 
     \Big) \, \psi \, , 
\ee
which by use of (\ref{theta_transf}) becomes (\ref{oddKP_psi_t5}).

\subsection{The commutative case}
If $\cA$ is commutative, imposing the \emph{reduction} condition (\ref{theta=kphi_t1}), i.e.
$\tilde{\theta} = \theta = k \, \phi_{t_1}$ with a constant $k$, the linear system (\ref{fun_oddKP}) 
takes the form
\be
    \frac{1}{\lambda} (\psi_{2[\lambda]} - \psi) 
  = F(\lambda)^{k-\frac{1}{2}} \Big(F(\lambda)^{-k-\frac{1}{2}} (\psi_{2[\lambda]}+\psi) \Big)_{t_1} \; .
                  \label{c-fun_oddKP-reduction}
\ee
Hence
\be
    \frac{1}{\lambda} \, (\psi_{2[\lambda]}-\psi) = \left\{ 
          \begin{array}{l@{\quad}c@{\quad}l@{\quad}l}
           F(\lambda)^{-1}(\psi_{2[\lambda]}+\psi)_{t_1}  &  &  k=-\frac{1}{2} & \mathrm{(BKP)} \vspace{.2cm} \\ 
           \left( F(\lambda)^{-1}(\psi_{2[\lambda]}+\psi) \right)_{t_1}  & \mbox{for} &  k=\frac{1}{2} & \mathrm{(BKP)} \\
           F(\lambda)^{-\frac{1}{2}} \left( F(\lambda)^{-\frac{1}{2}} (\psi_{2[\lambda]}+\psi) \right)_{t_1}
                                             &  & k=0 & \mathrm{(CKP)}
           \end{array} 
                                            \right. 
\ee
The CKP functional linear equation is half of the sum of the two BKP functional linear equations. 
In the remainder of this section we consider the case where $\phi$ is a $\mathbb{C}$-valued function and write
\be
       \phi = (\ln \tau^2)_{t_1} = 2 \, (\ln \tau)_{t_1} \, ,   \label{phi-tau}
\ee
with a function $\tau$. (In sections~\ref{subsec:BCKP_case1} and \ref{subsec:BCKP_case2} we 
use a different function $\tau$ given by $\phi = (\ln \tau)_{t_1}$.)

\begin{lemma}
The bilinear identity (\ref{bilinear-id}) with the reduction $\theta = k \, \phi_{t_1}$ 
implies
\be
    w(\lambda^{-1}) = \frac{\tau_{-2[\lambda]}}{\tau} \, F(-\lambda)^{k+\frac{1}{2}} \, , \qquad
    \tw(\lambda^{-1}) = \frac{\tau_{2[\lambda]}}{\tau} \, F(\lambda)^{-k+\frac{1}{2}} \, ,
        \label{c-w-reduction}
\ee
where $F(\lambda)$ now takes the form 
\be
    F(\lambda) = 1 - \lambda \, \Big(\ln \frac{\tau_{2[\lambda]}}{\tau} \Big)_{t_1}  \; .
\ee
\end{lemma}
\noindent
\textit{Proof:} We refer to some consequences of (\ref{bilinear-id}) derived in Appendix~A. 
(\ref{wtw-F}) can be written as
\bez
    \tw(\lambda^{-1}) = \frac{F(\lambda)}{w_{2[\lambda]}(\lambda^{-1})} \; . 
\eez
 From (\ref{wtw1}) we get
\bez
    w'_{2[\lambda]}(\lambda^{-1}) \, \tw(\lambda^{-1}) = \frac{1}{\lambda} \, (F(\lambda)-1) \, F(\lambda) 
      + \frac{1}{2} \, F'(\lambda) - \frac{\lambda}{2} \, (\theta_{2[\lambda]} - \theta) \, , 
\eez
where a prime indicates a partial derivative with respect to $t_1$, and thus 
\bez
     (\ln w_{2[\lambda]}(\lambda^{-1}))' = - \frac{1}{2} (\phi_{2[\lambda]}-\phi)
       + \frac{1}{2} (\ln F(\lambda))' - \frac{\lambda}{2 F(\lambda)} (\theta_{2[\lambda]} - \theta) \; .
\eez
Using (\ref{phi-tau}), the preceding equation can be integrated to
\bez
    w_{2[\lambda]}(\lambda^{-1}) = \frac{\tau}{\tau_{2[\lambda]}} \, F(\lambda)^{k+\frac{1}{2}}  \, ,
\eez
which is equivalent to the first equation in (\ref{c-w-reduction}). 
With its help, the equation we started with becomes the second of (\ref{c-w-reduction}).
\hfill $\square$
\vskip.2cm

By use of the lemma, and setting $z=\lambda^{-1}$, we find
\be
    \tw(z) = \left\{ \begin{array}{l@{\quad}c@{\quad}l@{\quad}l}
           w(-z)-z^{-1} \, w(-z)_{t_1}    &            &  k=-\frac{1}{2} & \mathrm{(BKP)} \\
           \tw(-z) + z^{-1} \tw(-z)_{t_1} & \mbox{for} &  k=\frac{1}{2}  & \mathrm{(BKP)} \\
           w(-z)                          &            &  k=0            & \mathrm{(CKP)}
           \end{array} \right. 
\ee
and thus the following relations for the Baker-Akhiezer function $\psi$ and its adjoint $\tpsi$, 
\be
   \left.  \begin{array}{l@{\quad}c@{\quad}l@{\quad}l}
           \tpsi(z) = -z^{-1} \psi(-z)_{t_1} &            &  k=-\frac{1}{2} & \mathrm{(BKP)} \\
           \psi(z) = z^{-1} \tpsi(-z)_{t_1}  & \mbox{for} &  k=\frac{1}{2}  & \mathrm{(BKP)} \\
           \tpsi(z) = \psi(-z)               &            &  k=0            & \mathrm{(CKP)}
           \end{array} \right. 
\ee

\begin{proposition}
The bilinear identity (\ref{bilinear-id}) with the reduction $\theta = k \, \phi_{t_1}$ implies the 
``differential Fay identity''
\be
   && \frac{\lambda+\mu}{\lambda-\mu} \, \tau_{2[\lambda]} \, \tau_{2[\mu]} \, 
      \Big( \lambda \, F_{2[\lambda]}(\mu)^{k+\frac{1}{2}} \, F(\mu)^{-k+\frac{1}{2}}
      - \mu \, F_{2[\mu]}(\lambda)^{k+\frac{1}{2}} \, F(\lambda)^{-k+\frac{1}{2}} \Big)  \nonumber \\
  &=& (\lambda+\mu) \, \tau \,\tau_{2[\lambda]+2[\mu]} - \lambda \mu \, \Big( (\tau_{2[\lambda]+2[\mu]})_{t_1} \, \tau
      - \tau_{2[\lambda]+2[\mu]} \, \tau_{t_1} \Big) \; .   \label{c-Fay-reduction}
\ee
\end{proposition}
\noindent
\textit{Proof:} This is obtained from (\ref{wtw2}) using (\ref{c-w-reduction}) and (\ref{Flamu}).
\hfill $\square$
\vskip.2cm

In the BKP case ($k = \pm 1/2$), the differential Fay identity (\ref{c-Fay-reduction}) is bilinear, 
\be
  \lefteqn{  (\lambda^{-1}+\mu^{-1}) (\tau_{2[\lambda]+2[\mu]} \, \tau - \tau_{2[\lambda]} \, \tau_{2[\mu]})
         }\hspace*{1.cm}&& \nonumber\\
    &=& (\tau_{2[\lambda]+2[\mu]})_{t_1} \, \tau - \tau_{2[\lambda]+2[\mu]} \, \tau_{t_1}
      + \frac{\lambda+\mu}{\lambda-\mu} \Big( (\tau_{2[\lambda]})_{t_1} \, \tau_{2[\mu]} 
          - \tau_{2[\lambda]} \, (\tau_{2[\mu]})_{t_1} \Big) \, ,
\ee
whereas in the CKP case ($k=0$) it is \emph{not}\footnote{This is in agreement with the fact that 
the CKP hierarchy cannot be expressed in Hirota bilinear form with a single $\tau$-function \cite{DJKM81TVI}.}, 
\be
    \lefteqn{ \frac{\lambda+\mu}{\lambda-\mu} \, \tau_{2[\lambda]} \, \tau_{2[\mu]}
    \Big( \lambda \, F_{2[\lambda]}(\mu)^{\frac{1}{2}} \, F(\mu)^{\frac{1}{2}}
    - \mu \, F_{2[\mu]}(\lambda)^{\frac{1}{2}} \, F(\lambda)^{\frac{1}{2}} \Big) }\hspace*{1.cm}&& \nonumber\\
    &=& (\lambda+\mu) \, \tau \, \tau_{2[\lambda]+2[\mu]} - \lambda \mu \, \Big( (\tau_{2[\lambda]+2[\mu]})_{t_1} \, \tau
    - \tau_{2[\lambda]+2[\mu]} \, \tau_{t_1} \Big) \; .
\ee
Expansion in powers of the indeterminates $\lambda$ and $\mu$ generates the BKP, respectively CKP, hierarchy 
equations.

\section{From a functional representation of the KP hierarchy to odd KP}
\label{sec:funct}
\setcounter{equation}{0}
A functional representation of the $m \times m$ matrix KP hierarchy is determined by \cite{DMH07Burgers}
\be
    \lambda^{-1} (\phi - \phi_{-[\lambda]}) - \phi_{t_1} 
    - (\phi - \phi_{-[\lambda]}) \, \phi 
  = \hat{\theta} - \hat{\theta}_{-[\lambda]}          \label{KP_func}
\ee
with an additional dependent variable $\hat{\theta}$, and 
$\phi_{[\lambda]}(\bt) = \phi(\bt + [\lambda])$ 
where $\bt = (t_1,t_2,t_3,\ldots)$ and $[\lambda] = (\lambda, \lambda^2/2, \lambda^3/3,\ldots)$. 
By expansion in powers of the indeterminate $\lambda$ 
and elimination of $\hat{\theta}$ one recovers the equations of the KP hierarchy. 
We note that although (\ref{KP_func}) contains a ``bare'' 
$\phi$ besides derivatives of it with respect to $t_n$, after elimination of $\hat{\theta}$ the 
resulting equations do not. Writing
\be
  \hat{\theta} = \tilde{\theta} - \frac{1}{2} ( \phi_{t_1} + \phi^2 ) \, , 
             \label{theta-ttheta-transf}
\ee
(\ref{KP_func}) takes the following form, after a Miwa shift, 
\be
    \lambda^{-1} (\phi_{[\lambda]} - \phi) - \frac{1}{2} ( \phi_{[\lambda]} + \phi)_{t_1} 
    -  \frac{1}{2} (\phi_{[\lambda]} - \phi)^2  
    + \frac{1}{2} \, [\phi , \phi_{[\lambda]} ] = \tilde{\theta}_{[\lambda]} - \tilde{\theta} 
   \; .                 \label{KP_func_alt}
\ee
Clearly, now one recovers the equations of the matrix KP hierarchy by expansion in 
powers of $\lambda$ and elimination of $\tilde{\theta}$. The first four equations from 
expansion of (\ref{KP_func_alt}) can be written as\footnote{Here we used e.g. the 
first equation to eliminate $\phi_{t_2}$ from the second. By use of (\ref{theta_transf}), 
(\ref{KP_phi_t2}) simplifies to $\phi_{t_2} = 2 \, \theta_{t_1}$ 
(see also remark~\ref{rem:oddKP->KP}). }
\be
   \phi_{t_2} &=& 2 \, \tilde{\theta}_{t_1} - [ \phi , \phi_{t_1} ] \, , \label{KP_phi_t2}  \\
   \tilde{\theta}_{t_2} &=& \frac{2}{3} \phi_{t_3} - \frac{1}{6} \phi_{t_1t_1t_1} 
       - (\phi_{t_1})^2 - \frac{1}{2} [ \phi , [ \phi , \phi_{t_1} ] ] 
       + [ \phi , \tilde{\theta}_{t_1} ]  \, ,     \label{KP_ttheta_t2} \\
   \phi_{t_4} &=& \frac{4}{3} \tilde{\theta}_{t_3} 
       + \frac{2}{3} \tilde{\theta}_{t_1t_1t_1} 
       + 2 \{ \phi_{t_1} , \tilde{\theta}_{t_1} \} 
       - \frac{1}{3} [ \phi , 2 \phi_{t_3} + \phi_{t_1t_1t_1} ]
       - [ \phi , (\phi_{t_1})^2 ] \, , \label{KP_phi_t4} \\
   \tilde{\theta}_{t_4} &=& \frac{4}{5} \phi_{t_5} - \frac{1}{3} \phi_{t_1t_1t_3} 
       + \frac{1}{30} \phi_{t_1t_1t_1t_1t_1} 
       - \frac{1}{6} \{ \phi_{t_1} , 4 \phi_{t_3} - \phi_{t_1t_1t_1} \} 
       + \frac{1}{2} \, (\phi_{t_1t_1})^2  \nonumber \\
    && - 2 \, (\tilde{\theta}_{t_1})^2
       - [ \phi_{t_1} , \tilde{\theta}_{t_1t_1} ] 
       + \frac{1}{3} [ \phi , 2 \tilde{\theta}_{t_3} + \tilde{\theta}_{t_1t_1t_1} ]
       - \frac{1}{6} [ \phi , [ \phi , 2 \phi_{t_3} + \phi_{t_1t_1t_1} ] ] \nonumber \\
    && + [ \phi , \{ \phi_{t_1} , \tilde{\theta}_{t_1} \} ]
       + \{ \tilde{\theta}_{t_1} , [  \phi , \phi_{t_1} ] \}
       + \frac{1}{2} [ \phi_{t_1} , [ \phi , \phi_{t_1t_1} ] ]
       - \frac{1}{2} [ \phi , \phi_{t_1} ]^2  \nonumber \\
    && - \frac{1}{2} [ \phi , [ \phi , (\phi_{t_1})^2 ] ]  \; . \label{KP_ttheta_t4}
\ee 
Solving the first equation for $\tilde{\theta}_{t_1}$ and using the resulting expression 
to eliminate $\tilde{\theta}$ from the second, results in the (potential) KP equation
\be
   4 \, \phi_{t_3} - \phi_{t_1t_1t_1} - 6 \, (\phi_{t_1})^2  
   - 3 \, \int \phi_{t_3t_3} \, \d t_1 + 6 \, \int [ \phi_{t_1} , \phi_{t_2} ] \, \d t_1 
   = 0 \; .
\ee
Instead of eliminating $\tilde{\theta}$ from (\ref{KP_func_alt}), which 
yields the matrix KP hierarchy, we can eliminate the derivatives of 
$\phi$ and $\tilde{\theta}$ with respect to the even-numbered variables, $t_{2n}$. 
This means we solve the equations resulting from (\ref{KP_func_alt}) 
for the derivatives of $\phi$ and $\tilde{\theta}$ with respect to $t_{2n}$, 
as in (\ref{KP_phi_t2}), (\ref{KP_ttheta_t2}), etc., compute their 
integrability conditions, and further use them to eliminate in the latter 
all derivatives with respect to even-numbered variables. In particular, 
$ \phi_{t_2t_4} = \phi_{t_4t_2} $ yields, after elimination of ``even'' 
derivatives, 
\be
 &&  \Big( 9 \, \phi_{t_5}  - 5 \, \phi_{t_1t_1t_3} + \phi_{t_1t_1t_1t_1t_1} 
       - \frac{15}{2} \{ \phi_{t_1} \, , \, \phi_{t_3} - \phi_{t_1t_1t_1} \}
       + 15 \, (\phi_{t_1})^3 
       + \frac{45}{4} \, \left( (\phi_{t_1t_1})^2 - 4 \, (\tilde{\theta}_{t_1})^2 \right)
       \nonumber \\
 && + \frac{45}{2} \, \{ \tilde{\theta}_{t_1} , [ \phi , \phi_{t_1} ] \}
    - \frac{15}{4} \, [ [  \phi , \phi_{t_1} ] , \phi_{t_1t_1} ]  
    - \frac{15}{2} \, [ [ \phi , \phi_{t_1t_1} ] , \phi_{t_1} ]
    -  \frac{45}{4} \, [ \phi , \phi_{t_1} ]^2     \Big)_{t_1}
    - 5 \, \phi_{t_3t_3}     \nonumber \\
 && + 15 \, \Big( [ \phi_{t_1} , \tilde{\theta}_{t_3} - \tilde{\theta}_{t_1t_1t_1} ] 
    + [ \tilde{\theta}_{t_1} , \phi_{t_3} + \frac{1}{2} \phi_{t_1t_1t_1} ] 
    + \frac{3}{2} \, [ \tilde{\theta}_{t_1t_1} \, , \, \phi_{t_1t_1} ] 
    - \frac{1}{2} [ \phi , [ \phi_{t_1} , \phi_{t_3} ]] \Big) 
    = 0 \, ,     \label{altoddKP1}
\ee
and from $ \tilde{\theta}_{t_2t_4} = \tilde{\theta}_{t_4t_2} $ one obtains 
another quite lengthy equation for the two dependent variables $\phi$ 
and $\tilde{\theta}$, involving only derivatives with respect to $t_1,t_3,t_5$. 
We verified independently with FORM \cite{Verm02} and Mathematica \cite{Wolf03} 
that via (\ref{theta_transf}) these two equations 
are equivalent to (\ref{oddKP1}) and (\ref{oddKP2}), which is our odd KP system.
\vskip.1cm

The structure displayed in (\ref{KP_phi_t2})-(\ref{KP_ttheta_t4}) in fact 
extends to the whole hierarchy, since the expansion of (\ref{KP_func_alt}) 
in powers of $\lambda$ has the following leading derivatives (which do not appear in 
the remaining terms, represented by dots), 
\be
   \lambda^{2n-1}: \quad 
   \frac{1}{2n} \, \phi_{t_{2n}} = \frac{1}{2n-1} \, \tilde{\theta}_{t_{2n-1}} + \ldots \, , 
    \qquad \quad
   \lambda^{2n}: \quad
   \frac{1}{2n} \, \tilde{\theta}_{t_{2n}} = \frac{1}{2n+1} \, \phi_{t_{2n+1}} + \ldots \, ,
\ee
where $n=1,2,\ldots$. Hence the method of computing the integrability conditions 
$\phi_{t_{2m}t_{2n}} = \phi_{t_{2n}t_{2m}}$ and 
$\tilde{\theta}_{t_{2m}t_{2n}} = \tilde{\theta}_{t_{2n}t_{2m}}$, and then eliminating 
all derivatives of $\phi$ and $\tilde{\theta}$ with respect to even-numbered variables, 
extends to the whole KP hierarchy. This yields a hierarchy of equations involving 
only derivatives with respect to odd-numbered variables and we have shown that its first 
member is our odd KP system. 
Because of the hierarchy property, it should then coincide with the odd KP hierarchy 
as formulated in section~\ref{subsec:GDS}, or generated by the linear system derived in 
section~\ref{sec:func_linsys}. 
\vskip.1cm

Above we started with a formulation of the KP hierarchy in terms of two dependent variables, $\phi$ 
and $\tilde{\theta}$ (or equivalently $\theta$). $\tilde{\theta}$ entered the stage as an 
auxiliary variable and its elimination leads to an expression for the KP hierarchy in terms 
of a single dependent variable, which is $\phi$. In this formulation of the KP hierarchy, 
the odd KP hierarchy is directly described as a subhierarchy (without further  
auxiliary variables as in the GDS formulation of section~\ref{subsec:GDS}). 
A particular consequence is that any method to construct exact solutions to the KP 
hierarchy in the formulation using the auxiliary dependent variable $\theta$ (or $\tilde{\theta}$) 
automatically yields solutions to the odd KP hierarchy. This fact will be used in 
section~\ref{sec:oddKPsol}. 
\vskip.1cm

We note that (\ref{KP_phi_t2}), (\ref{KP_ttheta_t2}), etc., are \emph{symmetries} of the odd KP 
hierarchy equations, with the help of which one recovers the whole KP hierarchy. 
\vskip.1cm

The next result will turn out to be crucial for establishing a relation between solutions to 
the (noncommutative) odd KP hierarchy and solutions to the BKP and CKP hierarchies. 
From now on we consider matrices over $\mathbb{R}$ or $\mathbb{C}$.

\begin{proposition} 
\label{prop:oddsymm}
The functional representation (\ref{KP_func_alt}) of the $m \times m$ matrix KP hierarchy is 
invariant under
\be
   \phi \mapsto \phi^\intercal \circ \varepsilon \, , \quad 
   \tilde{\theta} \mapsto - \tilde{\theta}^\intercal \circ \varepsilon  
    \, ,   \label{oddsymm}
\ee
where $\varepsilon(t_1,t_2,t_3,t_4,\ldots) := (t_1,-t_2,t_3,-t_4,\ldots)$, and $\phi^\intercal$ is 
the transpose of $\phi$. 
\end{proposition}
\noindent
\textit{Proof:} We consider (\ref{KP_func}) with $\phi$ and $\tilde{\theta}$ replaced by 
$\phi^\intercal \circ \varepsilon$ and $ - \tilde{\theta}^\intercal \circ \varepsilon$, respectively. 
Taking the transpose of the resulting equation,  
noting that $(f \circ \varepsilon)_{[\lambda]} = (f_{-[-\lambda]}) \circ \varepsilon$, and 
composing with $\varepsilon$ (which has the property $\varepsilon \circ \varepsilon = \mathrm{id}$), 
leads to
\bez
    \lambda^{-1}(\phi_{-[-\lambda]}-\phi) - \frac{1}{2} (\phi_{-[-\lambda]} + \phi)_{t_1} 
    - \frac{1}{2} (\phi_{-[-\lambda]}-\phi)^2 - \frac{1}{2} \, [\phi,\phi_{-[-\lambda]}] 
  = - \tilde{\theta}_{-[-\lambda]} + \tilde{\theta} \; .
\eez
With the substitution $\lambda \to -\lambda$ and a Miwa shift with $[\lambda]$, this becomes (\ref{KP_func}). 
\hfill $\square$
\vskip.2cm

As a consequence, the (matrix) KP hierarchy admits the symmetry reduction 
\be
    \phi = \phi^\intercal \circ \varepsilon \, , \quad 
   \tilde{\theta} = - \tilde{\theta}^\intercal \circ \varepsilon \;  .
\ee
Restricting to the \emph{odd} KP hierarchy, and setting $t_{2n} = 0$, $n=1,2,\ldots$, 
we have $\phi \circ \varepsilon = \phi$ and 
$\tilde{\theta} \circ \varepsilon = \tilde{\theta}$, hence the last conditions simplify to
\be
    \phi = \phi^\intercal \, , \quad 
   \tilde{\theta} = - \tilde{\theta}^\intercal  \;  .
        \label{oddKP_symm_red}
\ee
In particular, for $m=1$ we obtain $\tilde{\theta} = 0$, hence $\theta = 0$ by (\ref{theta_transf}), 
and thus the CKP hierarchy. The conditions (\ref{oddKP_symm_red}) are equivalent to those that 
determine the matrix CKP hierarchy, see section~\ref{subsec:GDS}. 

Obviously the reduction (\ref{oddKP_symm_red}) is \emph{not} 
compatible with the symmetries (the flows associated with $t_{2n}$) that extend the 
odd KP to the KP hierarchy. This example shows that a subhierarchy can admit a (symmetry) 
reduction that is \emph{not} a reduction of the complete hierarchy.

\begin{remark} 
A functional representation of the (noncommutative) \emph{discrete} KP hierarchy is given by \cite{DMH07Ricc}
\be
    \lambda^{-1} (\phi - \phi_{-[\lambda]}) - (\phi^+ - \phi_{-[\lambda]}) \, \phi = \hat{\theta}^+ - \hat{\theta}_{-[\lambda]} \, ,
    \label{dKP_func}
\ee
where $n \in \mathbb{Z}$ and $(\phi^+)_n = \phi_{n+1}$. To order $\lambda^0$, we obtain 
\be
    \phi_{t_1} - (\phi^+ - \phi) \, \phi = \hat{\theta}^+ - \hat{\theta} \; .     \label{dKP_la^0}
\ee
Subtracting this from (\ref{dKP_func}) yields (\ref{KP_func}), hence each $\phi_n$, $n \in \mathbb{Z}$, 
has to satisfy the KP hierarchy, thus also $\phi^+$.\footnote{By eliminating $\hat{\theta}$ and $\hat{\theta}^+$, 
one obtains the modified KP (mKP) hierarchy for $v$, where $v_{t_1} = \phi^+ - \phi$, and the Miura transformation.}
The transformation (\ref{theta-ttheta-transf}) converts the discrete KP hierarchy into
\be
   \lambda^{-1} (\phi_{[\lambda]} - \phi) - \frac{1}{2} (\phi_{[\lambda]} + \phi)_{t_1}
       - \frac{1}{2} (\phi_{[\lambda]} - \phi)^2 + \frac{1}{2} \, [\phi,\phi_{[\lambda]}] 
    &=&  \tilde{\theta}_{[\lambda]} - \tilde{\theta}  \, ,   \label{dKP_KP_tth}   \\
   \frac{1}{2} (\phi^+ + \phi)_{t_1} + \frac{1}{2} (\phi^+ - \phi)^2 + \frac{1}{2} \, [\phi,\phi^+]
          &=& \tilde{\theta}^+ - \tilde{\theta}  \; .   \label{dKP_la^0_tth} 
\ee
According to proposition~\ref{prop:oddsymm}, 
$\phi^+ = \phi^\intercal \circ \varepsilon$ and $\tilde{\theta}^+ = - \tilde{\theta}^\intercal \circ \varepsilon$   
solve (\ref{dKP_KP_tth}) if $\phi$ and $\tilde{\theta}$ do. 
Restricting the KP hierarchy (in the form presented in this section) to the \emph{odd} KP hierarchy, 
in the \emph{scalar} case ($m=1$) these conditions read
\be
     \phi^+ = \phi \, , \qquad
     \tilde{\theta}^+ = - \tilde{\theta} \, ,   \label{phi^+=phi,theta^+=-theta}
\ee
and (\ref{dKP_la^0_tth}) becomes $\theta = \tilde{\theta} = - \frac{1}{2} \phi_{t_1}$, 
which is the BKP reduction! We also refer to \cite{Jimbo+Miwa83} (p.~969) for a related result.  
\hfill $\square$
\end{remark}

\section{Solutions to the odd KP system and some of its reductions via a matrix Riccati system} 
\label{sec:oddKPsol}
\setcounter{equation}{0}
We consider the matrix linear system
\be
    Z_{t_n} = H^n \, Z \qquad \quad n=1,2,\ldots \, , \qquad 
    H = \left( \begin{array}{cc} R & Q \\ S & L \end{array} \right) \, , \qquad
    Z = \left( \begin{array}{c} X \\ Y \end{array} \right) \, , 
    \label{Z_tn=H^nZ}
\ee
where $L,Q,R,S$ are, respectively, constant $M \times M$, $N \times M$, 
$N \times N$ and $M \times N$ matrices over $\mathbb{C}$, $X$ is an 
$N \times N$ and $Y$ an $M \times N$ matrix. With suitable technical assumptions, 
the size of the matrices may also be infinite. 
The solution to the above linear system is given by
\be
    Z = \exp\Big( \xi(\bt,H) \Big) \, Z_0  \qquad \mbox{where} \qquad
    \xi(\bt,H) := \sum_{k=1}^\infty t_k \, H^k \; .
        \label{Z=exp(tH)Z_0}
\ee
For the new variable  
\be
      \Phi := Y \, X^{-1}  \, ,  \label{Phi=YX^-1}
\ee
assuming that $X$ possesses an inverse, (\ref{Z_tn=H^nZ}) implies the 
following hierarchy of matrix Riccati equations
\be
    \Phi_{t_n} = S_n + L_n \, \Phi - \Phi \, R_n - \Phi \, Q_n \, \Phi 
    \qquad \quad  n = 1,2,\ldots \, , 
          \label{Riccati}
\ee
where
\be
  \left( \begin{array}{cc} R_n & Q_n \\ S_n & L_n \end{array} \right) 
  := H^n 
\ee
(see \cite{DMH06nahier,DMH07Burgers,DMH07Ricc,DMH07Wronski}). Using its functional representation
\be
    \lambda^{-1} (\Phi - \Phi_{-[\lambda]}) = S + L \, \Phi - \Phi_{-[\lambda]} \, R 
   - \Phi_{-[\lambda]} \, Q \, \Phi \, , 
\ee
it turns out (see \cite{DMH07Burgers} for details) that $\Phi$ together with 
\be
    \hat{\Theta} = \Phi \, R   \label{Theta=PhiR}
\ee 
solves the $M \times N$ matrix KP$_Q$ hierarchy, which is determined by 
\be
    \lambda^{-1} (\Phi - \Phi_{-[\lambda]}) - \Phi_{t_1} 
    - (\Phi - \Phi_{-[\lambda]}) \, Q \, \Phi 
  = \hat{\Theta} - \hat{\Theta}_{-[\lambda]}   \; .       \label{KP_Q_func}
\ee
If $\mathrm{rank}(Q) = m$, hence  
\be
    Q = V \, U^\intercal   \label{Q=VU^t}
\ee
with an $M \times m$ matrix $U$ (with transpose $U^\intercal$) and an 
$N \times m$ matrix $V$, then 
\be
    \phi := U^\intercal \, \Phi \, V
\ee 
solves the $m \times m$ matrix KP hierarchy (\ref{KP_func}). 
By use of the first Riccati equation
\be
    \Phi_{t_1} = S + L \, \Phi - \Phi \, R - \Phi \, Q \, \Phi 
                 \label{Phi_t1}
\ee
in $\hat{\Theta} = \tilde{\Theta} - \frac{1}{2} ( \Phi_{t_1} + \Phi Q \Phi )$ 
(cf (\ref{theta-ttheta-transf})), and using (\ref{Theta=PhiR}), we obtain
\be
   \tilde{\Theta}  = \frac{1}{2} ( S + L \, \Phi + \Phi \, R ) \; . 
                     \label{tTheta_Riccati}
\ee 
Here we shall drop $S$ since it cancels out in $\tilde{\Theta}_{[\lambda]} - \tilde{\Theta}$. 
It follows that the $Q$-modified version of (\ref{KP_func_alt}) is satisfied 
as a consequence of the Riccati system. Recalling (\ref{theta_transf}), 
which now takes the form
\be
  \tilde{\Theta} = \Theta 
  + \frac{1}{2} \int ( \Phi \, Q \, \Phi_{t_1} - \Phi_{t_1} \, Q \, \Phi )
                    \; \d t_1 \, ,
\ee
we arrive at the following conclusion.

\begin{proposition}
\label{prop:oddKPsol}
Any solution $\Phi$ to the \emph{odd Riccati hierarchy}, i.e. the Riccati hierarchy (\ref{Riccati}) 
restricted to odd $n$, together with\footnote{By use of the Riccati system (\ref{Riccati}), this can 
also be written as $\Theta = \frac{1}{2} \int (S_2 + L_2 \Phi - \Phi R_2 - \Phi Q_2 \Phi) \, \d t_1$. 
The integrand is the right hand side of the Riccati equation for the variable $t_2$ (which, however, 
is prohibited in proposition~\ref{prop:oddKPsol}), so that $\Theta_{t_1} = \frac{1}{2} \, \Phi_{t_2}$, 
a symmetry of the odd KP (here odd KP$_Q$) hierarchy which we already met in remark~\ref{rem:oddKP->KP}. } 
\be
 \Theta = \frac{1}{2} \Big(L \, \Phi + \Phi \, R 
 - \int ( \Phi \, Q \, \Phi_{t_1} - \Phi_{t_1} \, Q \, \Phi ) \; \d t_1 \, \Big) 
    \, ,    \label{Theta-Phi}
\ee
solves the odd KP$_Q$ hierarchy.\footnote{Hence it solves in particular (\ref{oddKP1}) and 
(\ref{oddKP2}) with $\phi$ and $\theta$ replaced by matrices $\Phi$ and $\Theta$, and with the 
product modified by the constant matrix $Q$. } 
Furthermore, if (\ref{Q=VU^t}) holds, then 
\be
    \phi = U^\intercal \, \Phi \, V  \qquad \mbox{and} \qquad
    \theta = U^\intercal \, \Theta \, V     \label{phi,theta_mxm-oddKP}
\ee
solve the $m \times m$ matrix odd KP hierarchy (hence in particular the odd KP system 
(\ref{oddKP1}) and (\ref{oddKP2})). 
If $m=1$, then  
\be
    \phi = U^\intercal \, \Phi \, V  \qquad \mbox{and} \qquad
    \theta = \frac{1}{2} \, U^\intercal \, ( L \, \Phi + \Phi \, R ) \, V
    \label{phi,theta-Phi}
\ee
solve the scalar odd KP hierarchy (thus in particular (\ref{oddKP1c}) and (\ref{oddKP2c})).
\hfill $\square$
\end{proposition}

\begin{remark}
\label{rem:Ricc_red}
For some fixed $r \in \mathbb{N}$, $r>1$, let us impose the condition
\be
    H^r \, Z_0 = Z_0 \, P \, ,  \label{Riccati_r-reduction}
\ee
with an $N \times N$ matrix $P$, on the solution (\ref{Z=exp(tH)Z_0}) of the linear matrix system 
(\ref{Z_tn=H^nZ}). This implies $H^{n r} \, Z_0 = Z_0 \, P^n$ and thus $H^{n r} \, Z = Z \, P^n$ 
for $n \in \mathbb{N}$. 
Hence $R_{nr} X + Q_{nr} Y = X P^n$ and $S_{nr} X + L_{nr} Y = Y P^n$, which leads to the 
algebraic Riccati equations
\be
    S_{nr} + L_{nr} \Phi = Y P^n X^{-1} 
  = \Phi \, X P^n \, X^{-1}
  = \Phi \, ( R_{nr} + Q_{nr} \Phi )    \qquad \quad    n \in \mathbb{N}  \; .
\ee
The corresponding equations of the Riccati hierarchy then imply $\Phi_{t_{nr}} = 0$, for all 
$n \in \mathbb{N}$. The condition (\ref{Riccati_r-reduction}) thus ensures that $\Phi$ solves 
the $r$-reduction of the KP hierarchy ($r$th Gelfand-Dickey hierarchy). If $r$ is odd, this also 
yields a reduction of the odd KP hierarchy. Hence, adding the condition (\ref{Riccati_r-reduction}) 
to the assumptions of proposition~\ref{prop:oddKPsol}, (\ref{phi,theta_mxm-oddKP}) 
constitutes a solution to the $r$-reduction of the $m \times m$ matrix odd KP hierarchy. 
For $r=3$ this is the hierarchy with the pair (\ref{SK1}), (\ref{SK2}) as its first member, 
for $r=5$ it starts with the noncommutative generalization of the bSK and bKK equations, see 
section~\ref{subsec:red-oddKP}. 
\hfill $\square$
\end{remark}

In proposition~\ref{prop:oddKPsol} ``odd KP hierarchy'' more directly refers to the form  
in section~\ref{sec:funct}, where it has been described as a subhierarchy of the KP hierarchy, 
in the formulation of the latter involving the auxiliary variable $\theta$. 
In the scalar case, this hierarchy then admits reductions to the CKP and BKP hierarchy by imposing 
$\theta = 0$, respectively $\theta = - \frac{1}{2} \phi_{t_1}$ (see section~\ref{sec:oddKP}). 
In the following we show how the preceding proposition generates solutions to the BKP and the 
(matrix) CKP hierarchy.

\begin{lemma}
\label{lemma:Phi(anti)sym}
Let $M=N$. The transformation given by 
\be
   L \mapsto - R^\intercal \, , \quad 
   R \mapsto - L^\intercal \, , \quad 
   Q \mapsto \pm Q^\intercal  \, , \quad 
   S \mapsto \pm S^\intercal  \, , \quad 
   \Phi \mapsto \pm \Phi^\intercal \circ \varepsilon \, , 
   \label{Riccati-sym}
\ee
with $\varepsilon$ defined in proposition~\ref{prop:oddsymm}, 
leaves the Riccati hierarchy (\ref{Riccati}) invariant. 
\end{lemma}
\noindent
\textit{Proof:} The first four replacement rules in (\ref{Riccati-sym}) can be 
combined into
\be
    H \mapsto  - \mathcal{T} \, H^\intercal \, \mathcal{T}^{-1}  \qquad \mbox{with} \quad 
    \mathcal{T} = \left( \begin{array}{cc} 0 & \mp I_N \\ I_N & 0 \end{array} \right) \; .
    \label{Ht_BCKP}
\ee
This implies
\bez
   H^n \mapsto  (-1)^n \, \mathcal{T} (H^n)^\intercal \mathcal{T}^{-1} \, ,
\eez
and thus
\bez
    L_n \mapsto (-1)^n \, L_n^\intercal \, \quad 
    R_n \mapsto (-1)^n \, R_n^\intercal \, , \quad
    Q_n \mapsto \mp (-1)^n \, Q_n^\intercal \, , \quad
    S_n \mapsto \mp (-1)^n \,  S_n \; . 
\eez
Applying the map to the Riccati hierarchy (\ref{Riccati}), taking the transpose and 
using $(\Phi \circ \varepsilon)_{t_n} = (-1)^{n+1} \Phi_{t_n} \circ \varepsilon$, 
reproduces (\ref{Riccati}). 
\hfill $\square$
\vskip.2cm

As a consequence of the preceding lemma, we have the following symmetry reduction of the 
Riccati hierarchy (\ref{Riccati}),  
\be
   R = - L^\intercal \, , \qquad 
   Q^\intercal = \pm Q  \, , \qquad 
   S^\intercal = \pm S \, ,        \label{LRQS_pm_cond}
\ee
together with 
\be
    \Phi = \pm \Phi^\intercal \circ \varepsilon \; .      \label{Phi(anti)sym}
\ee
Restricting to the odd Riccati hierarchy, we are allowed to set $t_{2n} = 0$, $n=1,2,\ldots$. 
Then $\Phi$ given by (\ref{Phi=YX^-1}) solves the odd Riccati hierarchy and 
has the property $\Phi \circ \varepsilon = \Phi$. Furthermore, (\ref{LRQS_pm_cond}) and 
(\ref{Phi(anti)sym}), which now reads $\Phi = \pm \Phi^\intercal$, constitute a 
symmetry reduction of the odd Riccati hierarchy.

\begin{proposition}
\label{prop:matrixCKPsol}
Let $M=N$ and $\Phi$ a solution to the odd Riccati hierarchy with 
\be
    R = -L^\intercal \, , \qquad  S = S^\intercal \, , \qquad Q = V V^\intercal \, ,  
         \label{LRQSV_cond}
\ee
where $V$ is a constant $N \times m$ matrix. If 
\be
       \Phi^\intercal = \Phi  \, ,    \label{Phi^t=Phi}
\ee
then
\be
    \phi = V^\intercal \, \Phi \, V   \quad \mbox{and} \quad 
    \theta = V^\intercal \, \Theta \, V
\ee 
with $\Theta$ given in (\ref{Theta-Phi}) solve the $m \times m$ matrix CKP hierarchy (see 
section~\ref{subsec:GDS}).
\end{proposition}
\noindent
\textit{Proof:} The conditions (\ref{LRQSV_cond}) and (\ref{Phi^t=Phi}) correspond to the 
upper signs in (\ref{LRQS_pm_cond}). 
According to proposition~\ref{prop:oddKPsol}, $\phi$ and $\theta$ solve the $m \times m$ 
matrix odd KP hierarchy. Using (\ref{Theta-Phi}), (\ref{Phi^t=Phi}) and $Q^\intercal = Q$, 
one easily verifies that $\theta^\intercal = - \theta$ holds, which is the reduction to the 
matrix CKP hierarchy.
\hfill $\square$
\vskip.2cm

\begin{corollary}
\label{cor:CKPsol}
Let $M=N$ and $\Phi$ a solution to the odd Riccati hierarchy with (\ref{LRQSV_cond}), where $V$ 
is a constant $N$-component vector. If $\Phi^\intercal = \Phi$, then $\phi = V^\intercal \, \Phi \, V$ 
solves the CKP hierarchy.
\end{corollary}
\noindent
\textit{Proof:} The assertion follows from the last proposition, with $\Theta$ defined in (\ref{Theta-Phi}) 
and $m=1$, in which case the CKP reduction condition $\theta = 0$ holds. 
\hfill $\square$
\vskip.2cm

To obtain BKP solutions via proposition~\ref{prop:oddKPsol} is a bit less direct. 

\begin{proposition}
\label{prop:BKPsol}
Let $M=N$ and $\Phi$ a solution to the odd Riccati hierarchy subject to 
the conditions (\ref{LRQSV_cond}) with a constant $N$-component vector $V$. If $\Phi$ satisfies 
\be
    S + L \, \Phi + \Phi^\intercal \, L^\intercal - \Phi^\intercal \, Q \, \Phi = 0  \, , \label{Phit=Phi+}
\ee
then $\phi = V^\intercal \, \Phi \, V$ solves the BKP hierarchy. 
\end{proposition}
\noindent
\textit{Proof:} First we note that the fractional linear transformation 
$\Phi \mapsto \Phi^+ := (S + L \Phi)(R + Q \Phi)^{-1}$ 
(provided the inverse exists) leaves the Riccati hierarchy (\ref{Riccati}) invariant. 
This is so because this transformation is induced by $Z \mapsto H \, Z$, which leaves the linear 
matrix system (\ref{Z_tn=H^nZ}) invariant. We may then impose the symmetry reduction 
$\Phi^\intercal = \Phi^+$, i.e.  
\bez
     \Phi^\intercal = (S + L \Phi)(R + Q \Phi)^{-1} \, , 
\eez
which is (\ref{Phit=Phi+}). 
Using the definitions (\ref{phi,theta-Phi}) with $U=V$, the first Riccati equation (\ref{Phi_t1}), 
and then the last equation, we show that the BKP reduction condition is satisfied, 
\bez
     2 \, ( \theta + \frac{1}{2} \phi_{t_1} )
 &=& V^\intercal ( L \Phi - \Phi L^\intercal + \Phi_{t_1}) \, V   \\
 &=& V^\intercal ( S + 2 \, L \Phi - \Phi Q \Phi ) \, V   \\
 &=& V^\intercal ( L \Phi - \Phi^\intercal \, L^\intercal 
     + (\Phi^\intercal - \Phi) \, V  V^\intercal \, \Phi ) \, V   \\
 &=& V^\intercal L \Phi V - V^\intercal \Phi^\intercal \, L^\intercal V = 0 \; .
\eez
(One also finds $\phi^+ = \phi$ and $\theta^+ = - \theta$, cf (\ref{phi^+=phi,theta^+=-theta}).)
\hfill $\square$

\begin{remark}
The \emph{discrete} KP$_Q$ hierarchy is solved by 
a sequence $\Phi = (\Phi_n)_{n \in \mathbb{Z}}$ of solutions to the Riccati hierarchy (\ref{Riccati}) if 
$L \, \Phi - \Phi^+ \, R - \Phi^+ \, Q \, \Phi = 0$, where $\Phi_n^+ = \Phi_{n+1}$. This is the 
fractional linear transformation appearing in the proof of proposition~\ref{prop:BKPsol}, with $S=0$. 
It follows from (\ref{dKP_la^0}) by use of (\ref{Theta=PhiR}) and (\ref{Phi_t1}). 
\hfill $\square$
\end{remark}

\begin{remark}
The case with the lower signs in lemma~\ref{lemma:Phi(anti)sym} might be expected 
to be related to BKP. But it requires a skew-symmetric $Q$ and thus does not quite fit together with proposition~\ref{prop:oddKPsol}. However, writing $Q = \tilde{Q} L - L^\intercal \tilde{Q}$ with 
a rank one matrix $\tilde{Q} = V V^\intercal$, it turns out that 
$\phi = V^\intercal (L \Phi - \Phi L^\intercal) V = 2 \, V^\intercal L \Phi V $ 
solves the BKP equation (and its hierarchy), if $\Phi$ satisfies the conditions of 
lemma~\ref{lemma:Phi(anti)sym} with the lower signs. We shall elaborate on the underlying structure 
elsewhere. 
\hfill $\square$
\end{remark}

\begin{remark}
\label{rem:Z-BCKP}
As a consequence of (\ref{Z_tn=H^nZ}) and (\ref{LRQS_pm_cond}), 
which implies $H = - \mathcal{T} H^\intercal \mathcal{T}^{-1}$ with $\mathcal{T}$ defined in 
(\ref{Ht_BCKP}), we have
\be
    (Z^\intercal \, \mathcal{T} H^k \, Z)_{t_n} = 0  \qquad \mbox{for all odd $n$}  
\ee
and $k=0,1,\ldots$. Choosing $\mathcal{T}$ with the \emph{minus} sign, our CKP condition 
$\Phi^\intercal = \Phi$ originates from $Z^\intercal \, \mathcal{T} \, Z = 0$, and the 
BKP condition (\ref{Phit=Phi+}) corresponds to $Z^\intercal \, \mathcal{T} H \, Z = 0$. 
These conditions are the first two in a sequence that offers additional possibilities, 
\be
    Z^\intercal \, \mathcal{T} \, H^k \, Z = 0  \qquad \quad k=0,1,2,\ldots \; . \label{ZTH^kZ}
\ee 
We note that $(\mathcal{T} \, H^k)^\intercal = (-1)^{k+1} \mathcal{T} \, H^k$, so that 
the left hand side of (\ref{ZTH^kZ}) is a symmetric bilinear form if $k$ is odd, and 
skew-symmetric if $k$ is even. 
Invariance under a transformation $Z \mapsto G Z$, with a constant invertible matrix $G$,
requires $G^\intercal \mathcal{T} H^k G = \mathcal{T} H^k$. If the bilinear form is 
\emph{non-degenerate}\footnote{In the CKP case ($k=0$) this is fulfilled. In the BKP case 
(and more generally for  $k>0$), and if $S=0$ and $R=-L^\intercal$, which is the case we 
address in more detail below, the bilinear form is non-degenerate iff $\det(L) \neq 0$.}, 
this means that $G$ has to be (complex) orthogonal if $k$ is odd, and symplectic if $k$ is even. 
This connects with original work like \cite{DJKM81TVI}. It should be noticed, however, that 
the above method to construct solutions to the BKP hierarchy also works if the 
bilinear form is \emph{degenerate}. 
\hfill $\square$
\end{remark}

\begin{remark}
\label{rem:Ricc_BCKP_red}
Adding the $r$-reduction condition (\ref{Riccati_r-reduction}) to the assumptions of 
corollary~\ref{cor:CKPsol}, respectively proposition~\ref{prop:BKPsol}, they generate 
solutions to the $r$-reduction of the CKP, respectively BKP, hierarchy. 
For $r=3$, this yields solutions to the Kaup-Kupershmidt, respectively the Sawada-Kotera equation. 
For $r=5$, we obtain solutions to the bKK, respectively the bSK equation (see section~\ref{subsec:BKP&CKP}). 
We will not elaborate this further in this work, but a comparison with the results in  
\cite{Dye+Park01,Dye+Park02,Park+Dye02,HCR06,HCR07} would certainly be of interest. 
\hfill $\square$
\end{remark}

In the following subsections we elaborate some classes of solutions more explicitly. We 
consider the odd Riccati hierarchy with $M=N$, 
impose the conditions (\ref{LRQSV_cond}) with $S=0$, and treat the rank one case where 
$Q = V V^\intercal$ with a vector $V$. The choices (\ref{Q=-L^tK-KL}) and 
(\ref{Q=I+[L,K]}) below have their origin in certain normal forms of the matrix $H$, 
see \cite{DMH07Ricc}.

\subsection{A class of BKP and CKP solutions} 
\label{subsec:BCKP_case1}
Setting 
\be
    Q = R \, K - K \, L = - (L^\intercal \, K + K \, L)   \label{Q=-L^tK-KL}
\ee
with a \emph{symmetric} matrix $K$ (i.e. $K^\intercal = K$), (\ref{Z=exp(tH)Z_0}) 
can be computed explicitly (cf \cite{DMH07Ricc}) and we find the following solution to the 
odd Riccati hierarchy, 
\be
    \Phi = e^{\txi(\tbt,L)} \, \Phi_0 \, \left( e^{-\txi(\tbt,L^\intercal)} \,
           ( I_N + K \, \Phi_0 ) - K \, e^{\txi(\tbt,L)} \, \Phi_0 \right)^{-1} \, , 
\ee
where $\Phi_0 = Y_0 \,  X_0^{-1}$ and 
\be
     \txi(\tbt,L) = \sum_{k=0}^\infty t_{2k+1} \, L^{2k+1} \; .
\ee 
Assuming $\Phi_0$ invertible, this simplifies to
\be
      \Phi = \left( e^{-\txi(\tbt,L^\intercal)} \,
           ( \Phi_0^{-1} + K ) \, e^{-\txi(\tbt,L)} - K \right)^{-1} \; . 
             \label{Phi_solution1}
\ee
Using $Q = V V^\intercal$, the cyclicity of the trace, and $\tr \ln = \ln \det$, we obtain 
\be
 \phi &=& V^\intercal \Phi V = \tr( Q \, \Phi) = - \tr( (L^\intercal \, K + K \, L) \, \Phi ) \nonumber \\
      &=& (\ln \tau)_{t_1} \quad \mbox{with} \quad 
 \tau = \det( \Phi_0^{-1} + K - e^{\txi(\tbt,L^\intercal)} \, K \, e^{\txi(\tbt,L)} )
        \; .    \label{phi_solution1}
\ee 
Here $K,L,V$ have to solve the rank one condition 
\be
     L^\intercal \, K + K \, L = - V V^\intercal  \; .  \label{rank(L^tK+KL)=1}
\ee
In order that (\ref{phi_solution1}) solves the CKP or the BKP hierarchy, (\ref{Phi^t=Phi}), respectively (\ref{Phit=Phi+}), still has to be satisfied. 
\vskip.1cm

\noindent
\textbf{CKP.} If $\Phi_0$ is symmetric, i.e. $\Phi_0^\intercal = \Phi_0$, then also $\Phi$ given 
in (\ref{Phi_solution1}). 
We can thus express $\tau$ in (\ref{phi_solution1}) as 
\be
  \tau = \det( C - e^{\txi(\tbt,L^\intercal)} \, K \, e^{\txi(\tbt,L)} ) 
          \label{case1_CKP_sol}
\ee
with an arbitrary constant \emph{symmetric} $N \times N$ matrix $C$, i.e. $C^\intercal = C$.
According to corollary~\ref{cor:CKPsol}, this determines a solution $\phi = (\ln \tau)_{t_1}$ to the 
CKP hierarchy, provided that $K$ and $L$ satisfy (\ref{rank(L^tK+KL)=1}). 
\vskip.1cm

\noindent
\textbf{BKP.} We have to elaborate the BKP condition (\ref{Phit=Phi+}) (with $S=0$). Using 
(\ref{Q=-L^tK-KL}), it can be expressed as 
\be
   L^\intercal \, ( \Phi^{-1} + K ) = - ( \Phi^{-1} + K )^\intercal \, L \; . 
\ee
Inserting (\ref{Phi_solution1}), written in the form
\be
    \Phi^{-1} + K = e^{-\txi(\tbt,L^\intercal)} \,
           ( \Phi_0^{-1} + K ) \, e^{-\txi(\tbt,L)} \, , 
\ee
this reduces to
\be
 \frac{1}{2} \, C := L^\intercal \, ( \Phi_0^{-1} + K ) 
                   = - ( \Phi_0^{-1} + K )^\intercal \, L 
                   = - \frac{1}{2} \, C^\intercal \, , 
\ee
i.e. $C$ has to be a \emph{skew-symmetric} matrix. 

It is known that BKP $\tau$-functions can be expressed as the square of a Pfaffian. In the following we 
translate (\ref{phi_solution1}) into such a form, assuming that $L$ is invertible.
We may replace $\tau$ given in (\ref{phi_solution1}) by 
\be
 \tau = \det( C - 2 \, e^{\txi(\tbt,L^\intercal)} \, L^\intercal  K \, e^{\txi(\tbt,L)} ) \, ,
\ee 
since the two expressions differ only by a constant factor that drops out in $\phi=(\ln \tau)_{t_1}$.
Using $L^\intercal K = - K \, L - V V^\intercal$, this becomes
\be
 \tau = \det(e^{\txi(\tbt,L)})^2 \, 
        \det( A + V V^\intercal ) 
 \qquad \mbox{where} \quad
 A := e^{-\txi(\tbt,L^\intercal)} \, C \, e^{-\txi(\tbt,L)} - L^\intercal K + K \, L 
       \; .   \label{BKP_tau_VV^t}
\ee 
If the size $N$ of the matrices is even, then $\det(A + V V^\intercal) = \det(A)$ (for skew-symmetric $A$, 
see e.g. (2.92) in \cite{Hiro04}) leads to
\be
    \tau = \det\left( C - e^{\txi(\tbt,L^\intercal)} \, ( L^\intercal K - K \, L ) \, 
           e^{\txi(\tbt,L)} \right) \; .   \label{case1_BKP_tau}
\ee
This is the determinant of a skew-symmetric matrix, hence $\tau$ 
can be expressed as the square of the Pfaffian of this matrix. 
If $N$ is odd, then $\det(A)=0$, but (\ref{BKP_tau_VV^t}) with a suitable choice of $V$ can still lead 
to non-trivial solutions. In this case we can use the identity
\be
    \det( A + V V^\intercal ) = \det\left( \begin{array}{cc} 0 & V^\intercal \\
                                                            -V & A \end{array} \right)
   = \left( \mathrm{Pf}\left( \begin{array}{cc} 0 & V^\intercal \\
                                                            -V & A \end{array} \right) \right)^2
\ee
(see Appendix~B) to express $\tau$ as the square of a Pfaffian.\footnote{The factor 
$\det(e^{\txi(\tbt,L)})^2$ in (\ref{BKP_tau_VV^t}) can be dropped since it does not 
influence $\phi_{t_n}$.} 
\vskip.2cm

A subclass of solutions is obtained by choosing
\be
       L = \mathrm{diag}(p_1,\ldots,p_N)   \label{L_diag}
\ee
with constants $p_i$, $i=1,\ldots,N$. The solution to (\ref{rank(L^tK+KL)=1}) is then given by 
\be
    K_{ij} = - \frac{v_i v_j}{p_i + p_j}  \qquad \quad i,j = 1, \ldots, N \, , 
\ee
assuming $p_i + p_j \neq 0$ for all $i,j$ and writing $V^\intercal =(v_1,\ldots,v_N)$. 
From this one recovers in particular BKP and CKP multi-soliton solutions (see also 
\cite{DKM81TII,DJKM81TVI,Jimbo+Miwa83} for different approaches).

\subsubsection{Examples}

\begin{example}
\label{ex:case1_CKP}
We consider the CKP case with (\ref{L_diag}). For $N=1$, (\ref{case1_CKP_sol}) becomes 
$\tau = 1 + b \, e^{2 \, \txi(\tbt,p)}$ with 
$b = \frac{v^2}{2 c p}$, dropping an irrelevant factor $c$. This yields a regular 
solution if $b>0$, and $u = \phi_{t_1}$ then describes a single line soliton. 
For $N = 2$ and $C = \mathrm{diag}(c_1,c_2)$ we obtain, dropping an irrelevant factor $c_1c_2$, 
\be
   \tau = 1 + b_1 \, e^{2 \, \txi(\tbt,p_1)}
         + b_2 \, e^{2 \, \txi(\tbt,p_2)} 
         + b_1 \, b_2 \, 
           \left(\frac{p_1-p_2}{p_1+p_2}\right)^2 \, 
           e^{ 2 \, \txi(\tbt,p_1) + 2 \, \txi(\tbt,p_2) } \, , 
  \quad
  b_i := \frac{v_i^2}{2 c_i p_i} \, , 
\ee
assuming $p_1,p_2,c_1,c_2 \neq 0$ and $p_2 \neq - p_1$. 
If the parameters are real and such that $b_1,b_2>0$, this yields a regular 
CKP solution $\phi$, and $u = \phi_{t_1}$ generically describes two oblique line solitons. 
In this case we can simplify the above expression by writing $b_i = \exp(2 a_i)$ with constants 
$a_i$, $i=1,2$. 
\hfill $\square$
\end{example}

\begin{example}
\label{ex:case1_BKP} 
In the BKP case with (\ref{L_diag}), we consider $N=2$, hence
\be
 L = \left( \begin{array}{cc} p_1 & 0 \\ 0 & p_2 \end{array} \right) \, , \qquad
 C = \left( \begin{array}{cc} 0 & c \\ -c & 0 \end{array} \right) \, ,  \qquad
 V = \left( \begin{array}{c} v_1 \\ v_2 \end{array} \right) \; . 
       \label{L2x2diagCV_BKP}
\ee
(\ref{case1_BKP_tau}) leads to $\tau = \mathfrak{p}^2$ with 
\be
   \mathfrak{p} = c + v_1 v_2 \, \frac{p_1-p_2}{p_1+p_2} \, 
     e^{ \txi(\tbt,p_1) + \txi(\tbt,p_2)} \, , 
          \label{BKP_1line_soliton_Pfaffian}
\ee
if $p_1+p_2 \neq 0$. Without restriction of generality we can set $v_1 = v_2 =1$. 
For real $c,p_1,p_2$, the function $\phi$ is then 
regular (for all $t_1,t_2,\ldots$) iff $c \, (p_1^2-p_2^2) > 0$,  
and $u = \phi_{t_1}$ describes a single line soliton. 
\hfill $\square$
\end{example}
\vskip.2cm

Solutions can be superposed as follows. If 
$(L_i,V_i,K_i,C_i)$, $i=1,2$, are two sets of matrix data that determine (BKP or CKP)  
solutions, then 
\be
 && L = \left( \begin{array}{cc} L_1 & 0 \\ 
                               0   & L_2
             \end{array} \right)  \, , \quad
    V = \left( \begin{array}{c} V_1 \\ V_2
             \end{array} \right) \, , \quad 
    K = \left( \begin{array}{cc} K_1 & K_{12} \\ 
                          K_{12}^\intercal & K_2
             \end{array} \right) \, , \quad 
    C = \left( \begin{array}{cc} C_1 & 0 \\ 
                                 0   & C_2
             \end{array} \right)    \label{LVKC_super}
\ee
determine a new solution, provided that a solution $K_{12}$ exists to
\be
   L_1^\intercal \, K_{12} + K_{12} \, L_2 = - V_1 \, V_2^\intercal \; .
          \label{case1_super_cond}
\ee

\begin{figure}[t] 
\begin{center} 
\resizebox{6.cm}{!}{
\includegraphics{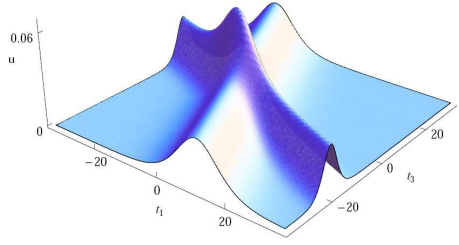}
}
\end{center} 
\caption{Plot of $u = \phi_{t_1} = 2 \, (\ln|\mathfrak{p}|)_{t_1t_1}$ (BKP) at $t_5,t_7, \ldots =0$ 
with $\mathfrak{p}$ given by (\ref{BKP_L2x2diag_super}), 
where $p_1 = 1/2, p_2=-1/4, p_3=1, p_4=-3/4$ and $c_1= c_2 =1$. 
\label{fig:BKP_2line_soliton} }
\end{figure}

\begin{figure}[t] 
\begin{center} 
\resizebox{6.cm}{!}{
\includegraphics{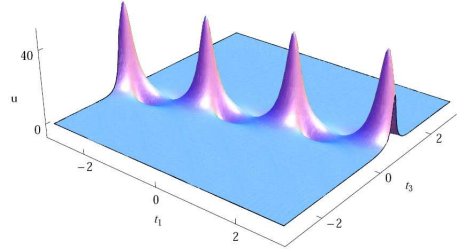}
}
\end{center} 
\caption{Plot of $u = \phi_{t_1} = 2 \, (\ln |\mathfrak{p}|)_{t_1t_1}$ at $t_5,t_7, \ldots =0$  
with $\mathfrak{p}$ given by (\ref{BKP_Pf_cL2x2diag_super}), 
where $p_1 = 1+3 \mathrm{i}/2$, $p_2=1+\mathrm{i}$ and $c=1$. 
The array of BKP solitons extends periodically to infinity.
\label{fig:BKP_periodic} }
\end{figure}

\begin{example}
\label{ex:case1_BKP_L4x4diag_conj_super}
We consider the BKP case. By superposition of two solutions of the form given in example~\ref{ex:case1_BKP}, 
setting $V_1=V_2=(1,1)^\intercal$, one obtains $\tau = \mathfrak{p}^2$ with 
\be
  \mathfrak{p} 
  = b \, \Big( \tilde{c}_1 \tilde{c}_2 + \tilde{c}_1 \, e^{\txi(\tbt,p_3) + \txi(\tbt,p_4) } 
 + \tilde{c}_2 \, e^{\txi(\tbt,p_1) + \txi(\tbt,p_1) } 
 + a \, e^{ \txi(\tbt,p_1) + \txi(\tbt,p_2) + \txi(\tbt,p_3) + \txi(\tbt,p_4) } 
     \Big)  \, ,    \label{BKP_L2x2diag_super}
\ee
where
\be
    a = \frac{(p_1-p_3)(p_2-p_3)(p_1-p_4)(p_2-p_4)}{(p_1+p_3)(p_2+p_3)(p_1+p_4)(p_2+p_4)} \, , \qquad
    b = \frac{(p_1-p_2)(p_3-p_4)}{(p_1+p_2)(p_3+p_4)} \, , 
\ee
and $\tilde{c}_1 = c_1 (p_1+p_2)/(p_1-p_2)$, $\tilde{c}_2 = c_2 (p_3+p_4)/(p_3-p_4)$. 
If $p_1 \neq p_2$ and $p_3 \neq p_4$, we may drop 
the factor $b$. With real parameters and $\tilde{c_1}, \tilde{c_2} >0$, one recovers a well-known expression 
for the 2-soliton solution ($a>0$) to the BKP hierarchy \cite{Hiro86,Hiro04}, see also Fig.~\ref{fig:BKP_2line_soliton}. 
Allowing the parameters to be complex, we can superpose the solution data 
(\ref{L2x2diagCV_BKP}) and the complex conjugate data, so that 
\be
  \mathfrak{p} = \Big| c + \frac{p_1-p_2}{p_1+p_2} \, e^{\txi(\tbt,p_1)
   + \txi(\tbt,p_2) } \Big|^2
   + \Big( \frac{\mathrm{Im}(p_1) \, \mathrm{Im}(p_2)}{ \mathrm{Re}(p_1) \, \mathrm{Re}(p_2) } 
     - \Big| \frac{p_1-p_2^\ast}{p_1+p_2^\ast} \Big|^2  \Big) \, 
   e^{2 \, \mathrm{Re}( \txi(\tbt,p_1) + \txi(\tbt,p_2) )} 
    \, . \quad
   \label{BKP_Pf_cL2x2diag_super}
\ee
A regular solution from this family is plotted in Fig.~\ref{fig:BKP_periodic}. 
See also Appendix~C for a general receipe to obtain real solutions from complex matrix data.  
\hfill $\square$
\end{example}

\subsection{Another class of BKP and CKP solutions} 
\label{subsec:BCKP_case2}
Now we set $R=L$, so that $L$ is skew-symmetric, i.e. $L^\intercal = -L$, and 
\be
    Q = I_N + [L,K]  \label{Q=I+[L,K]}
\ee
with a symmetric matrix $K$. 
Assuming $\Phi_0$ invertible, computation of (\ref{Z=exp(tH)Z_0}) 
(cf \cite{DMH07Ricc}) leads to the following solution to the odd Riccati hierarchy,
\be
   \Phi = \left( e^{ \txi(\tbt,L) } \, ( \Phi_0^{-1} + K ) \, e^{ -\txi(\tbt,L) } 
          + \txi'(\tbt,L) - K  \right)^{-1} \, ,   \label{Phi_solution2}
\ee
where 
\be
    \txi'(\tbt,L) := \sum_{n=0}^\infty (2n+1) \, t_{2n+1} \, L^{2n}  \; .
\ee
If also $Q = V V^\intercal$ with a vector $V$, hence $K,L,V$ have to satisfy the rank one condition 
\be
    I_N + [L,K] = V V^\intercal \, ,  \label{I+[L,K]=VV^t}
\ee 
then we obtain 
\be
 \phi &=& V^\intercal \Phi V = \tr ( (I_N + [L,K]) \, \Phi )  \nonumber \\
      &=& (\ln \tau)_{t_1}  \qquad \mbox{with} \quad  
 \tau = \det\Big( e^{\txi(\tbt,L)} \, (\Phi_0^{-1} + K) \, e^{-\txi(\tbt,L)} 
          + \txi'(\tbt,L) - K \Big)  \;  . \label{phi_solution2}
\ee 
In order that (\ref{phi_solution2}) solves the CKP or the BKP hierarchy, the condition (\ref{Phi^t=Phi}), 
respectively (\ref{Phit=Phi+}), still has to be elaborated. 
\vskip.1cm

\noindent
\textbf{CKP.} If $\Phi_0$ is symmetric, then also $\Phi$. We can then replace the above 
function $\tau$ by 
\be 
   \tau = \det\Big( e^{\txi(\tbt,L)} \, C \, e^{-\txi(\tbt,L)} 
          + \txi'(\tbt,L) - K \Big)  \, , \label{case2_CKP_tau}
\ee 
with an arbitrary constant \emph{symmetric} $N \times N$ matrix $C$. 
According to corollary~\ref{cor:CKPsol}, this determines a solution $\phi$  
to the CKP hierarchy, if $K$ and $L$ satisfy (\ref{I+[L,K]=VV^t}). 
\vskip.1cm

\noindent
\textbf{BKP.} The condition (\ref{Phit=Phi+}) (with $S=0$) can be written in the form
\be
   L \, ( \Phi^{-1} + K ) - ( \Phi^{-1} + K )^\intercal \, L = - I_N  \; .
\ee
Inserting (\ref{Phi_solution2}), rewritten as
\be
 \Phi^{-1} + K = e^{ \txi(\tbt,L) } \, ( \Phi_0^{-1} + K ) \, e^{ -\txi(\tbt,L) } 
          + \txi'(\tbt,L) \, , 
\ee
leads to
\be
   L \, ( \Phi_0^{-1} + K ) - ( \Phi_0^{-1} + K )^\intercal \, L = - I_N  \, ,
\ee
which is 
\be
    C^\intercal = - C \qquad \mbox{where} \qquad 
    C := 2 \, L \, ( \Phi_0^{-1} + K ) + I_N \, ,   \label{case2_BKP_C^t=-C}
\ee
i.e. $C$ has to be \emph{skew-symmetric}.
\vskip.1cm

Next we translate (\ref{phi_solution2}) in the BKP case into a form, 
where $\tau$ is the determinant of a skew-symmetric matrix, under the assumption that $\det(L) \neq 0$. 
According to remark~\ref{rem:Z-BCKP}, the latter condition corresponds to the \emph{genuine} BKP case. 
A function equivalent to $\tau$ given in (\ref{phi_solution2}) is then 
\be
   \tau &=& \det\Big( e^{\txi(\tbt,L)} \, ( C - I_N ) \, e^{-\txi(\tbt,L)} 
          + 2 L \, (\txi'(\tbt,L) - K) \Big)   \nonumber \\
        &=& \det( A - V V^\intercal )  \qquad \mbox{where} \quad
            A := e^{\txi(\tbt,L)} \, C \, e^{-\txi(\tbt,L)} 
                - (K L + L K) + 2 L \, \txi'(\tbt,L) \; . 
       \label{case2_BKP_tau_pre}
\ee
This is the determinant of the sum of the skew-symmetric matrix $A$ and a rank one matrix. 
If $N$ is even, then $\det( A - V V^\intercal ) = \det(A)$ and thus 
\be
   \tau = \det\Big( e^{\txi(\tbt,L)} \, C \, e^{-\txi(\tbt,L)} 
          - (K L + L K) + 2 L \, \txi'(\tbt,L) \Big) \, , 
          \label{case2_BKP_tau}
\ee
which is then the square of the Pfaffian of $A$.

\begin{remark}
(\ref{Q=I+[L,K]}) implies $\tr(L^k (Q-I_N)) = 0$, $k=0,1,\ldots,N-1$.\footnote{There 
are no independent equations for $k>N-1$ because of the Cayley-Hamilton theorem.} 
These constraints are obstructions to solving (\ref{Q=I+[L,K]}) for $K$ \cite{Brus+Ragn83}.
In particular, we have $\tr(Q) = N$. Hence $V$ lies on a sphere in $N$ dimensions. 
Since the (complex) orthogonal group acts transitively on a (complexified) 
sphere (see e.g. Lemma 4.1 in \cite{Mori98}), $V$ can be transformed to $V=(1,\ldots,1)^\intercal$. 
Since a similarity transformation of the matrices leaves (\ref{phi_solution2}) invariant, this 
means that without loss of generality we can set $V=(1,\ldots,1)^\intercal$, as long as 
no restrictions are placed on the antisymmetric matrix $L$.
\hfill $\square$
\end{remark}

Choosing $\Phi_0$ such that
\be
      [ \Phi_0^{-1} + K \, , \, L ] = 0  \, ,   \label{comm_L}
\ee
the above solutions become \emph{rational} functions of $t_1,t_3,t_5,\ldots$.\footnote{For other 
approaches to rational solutions see \cite{DJKM82TIV,Gils+Nimm90,Nimm+Orlo05} in the BKP and 
\cite{DJKM81TVI,DFL02,Dubr+Lisi02} in the CKP case. }
We confine ourselves to this case in the following examples. 
For the matrix $C$ (which has to be symmetric in the CKP and skew-symmetric in the BKP case), 
(\ref{comm_L}) implies $[ C \, , \, L ] = 0$.

\subsubsection{Examples}

\begin{example}
\label{ex:case2_N=2} 
Let $N=2$ and  
\be
  L = \left( \begin{array}{cc} 0 & p \\ -p & 0 \end{array} \right) 
\ee
with a constant $p$. According to the last remark we can set $V^\intercal = (1,1)$ 
without restriction of generality. The solution to (\ref{I+[L,K]=VV^t}) is then 
given by
\be
 K = c \, I_2 
 + \frac{1}{2p} \left( \begin{array}{cc} - 1 & 0 \\ 0 & 1 \end{array} \right)
     \label{case2_N=2_K}
\ee
with an arbitrary constant $c$. The condition (\ref{comm_L}) leads to
$C = a \, I_2 + b \, L$ with constants $a,b$. 

In the CKP case, $b=0$ and the resulting term in (\ref{phi_solution2}) 
involving $a$ can be absorbed by redefinition of $c$. We obtain
\be
   \tau = \eta^2 - \frac{1}{4 \, p^2} \qquad
   \mbox{where} \qquad  
     \eta(\tbt,p,c) 
   = \sum_{n=0}^\infty (-1)^n (2n+1) \, t_{2n+1} \, p^{2n} - c \; .  \label{eta}
\ee
If $p$ is imaginary, the corresponding CKP solution is real and regular. 
Treating $t_5$ as a `time' variable (and freezing the higher variables), 
$u = \phi_{t_1} = (\ln\tau)_{t_1t_1}$ describes a line soliton (with rational decay) 
moving in $t_1t_3$-space.  

In the BKP case, (\ref{case2_BKP_C^t=-C}) requires $b = \frac{1}{2} p^{-2}$, hence $C = 2a \, L$. 
In (\ref{case2_BKP_tau}), $a$ can be absorbed by redefinition of $c$. Hence we can set 
$C=0$ without loss of generality. We obtain 
$\mathrm{Pf}(A)= 2 p \, \eta(\tbt,p,c)$, 
which cannot provide us with a real and regular BKP solution. 
\hfill $\square$
\end{example}
\vskip.2cm

Given two sets of matrix data $(L_i,V_i,K_i)$, $i=1,2$, that determine (BKP or CKP) solutions, 
we can superpose them as follows, 
\be
     L = \left(\begin{array}{cc} L_1 & 0 \\0 & L_2 \end{array}\right) \, , \qquad
     V = \left(\begin{array}{c} V_1 \\ V_2 \end{array}\right) \, , \qquad
     K = \left(\begin{array}{cc} K_1 & K_{12} \\ K_{12}^\intercal & K_2 \end{array}\right) \; .
\ee
Then (\ref{I+[L,K]=VV^t}) is satisfied if $K_{12}$ solves 
\be
   L_1 \, K_{12} - K_{12} \, L_2 = V_1 \, V_2^\intercal \; .
          \label{case2_super}
\ee
\vskip.2cm

\begin{figure}[t] 
\begin{center} 
\resizebox{10cm}{!}{
\includegraphics{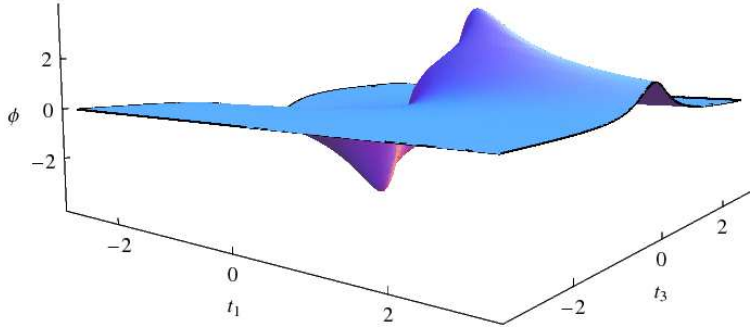}
}
\end{center} 
\caption{Plot of the CKP solution $\phi$ determined by (\ref{CKP_2x2super_tau}) 
at $t_5,t_7,\ldots =0$, with $p_2^\ast=p_1=1+\mathrm{i}$ and $c_1 = c_2 = 0$. 
This configuration simply moves in the $t_1t_3$-plane with 
varying $t_5$. 
\label{fig:CKP_2x2super} }
\end{figure}

\begin{figure}[t] 
\begin{center} 
\resizebox{10cm}{!}{
\includegraphics{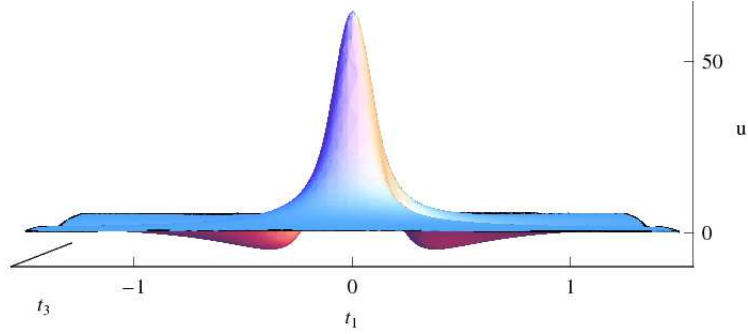}
}
\end{center} 
\caption{A lump solution to the BKP equation. Plot of $u =\phi_{t_1}$ at $t_5,t_7,\ldots =0$ 
for the solution given by (\ref{BKP1lump}) with $c=0$ 
and $p= 1 + 9 \mathrm{i}/10$.
   \label{fig:BKP1lump} }
\end{figure}

\begin{example}
\label{ex:case2_N=4}  
We superpose two solutions of the form given in example~\ref{ex:case2_N=2}. 
The solution to (\ref{case2_super}) is then
\be
    K_{12} = \left( \begin{array}{cc} 
             - \frac{1}{p_1+p_2} & - \frac{1}{p_1-p_2} \\
             \frac{1}{p_1-p_2} & \frac{1}{p_1+p_2}
                    \end{array} \right) \; .  \label{case2_N=4_K12}
\ee

In the CKP case, (\ref{phi_solution2}) with $C = 0$ yields
\be
 \tau = \Big( \eta_1 \, \eta_2 
  - \frac{1}{(p_1-p_2)^2} - \frac{1}{(p_1+p_2)^2} 
  - \frac{1}{4 p_1 p_2} \Big)^2
  - \frac{1}{4} \Big( \frac{\eta_1}{p_2} - \frac{\eta_2}{p_1} \Big)^2
  - \frac{1}{p_1 p_2 \, (p_1-p_2)^2} \, , \label{CKP_2x2super_tau}
\ee
where $\eta_i = \eta(\tbt,p_i,c_i)$, $i=1,2$ (see (\ref{eta})). 
If $p_2 = p_1^\ast$ (the complex conjugate of $p_1$), $c_2 = c_1^\ast$, and 
$\mathrm{Re}(p_1) \, \mathrm{Im}(p_1) \neq 0$, this expression is real 
(see also Appendix~C) 
and strictly positive, and thus determines a regular solution to  
the CKP hierarchy. See also Fig.~\ref{fig:CKP_2x2super}. 

In the BKP case, we obtain 
\be
  \mathrm{Pf}(A) = 4 p_1 p_2 \, \Big( \eta_1 \eta_2 
          - 2 \, \frac{p_1^2+p_2^2}{(p_1^2-p_2^2)^2} \Big)  \, ,
\ee
where again $\eta_i = \eta(\tbt,p_i,c_i)$, $i=1,2$. 
Choosing $p_2^\ast = p_1 =: p$ and $c_2^\ast = c_1 =:c$, this takes the form
\be
  \mathrm{Pf}(A) = 4 \, |p|^2 \Big( \frac{\mathrm{Re}(p^2)}{\mathrm{Im}(p^2)^2}
    + | \eta(\tbt,p,c) |^2 \Big)     \, ,    \label{BKP1lump}
\ee
which is strictly positive if $\mathrm{Re}(p^2)>0$, 
hence the solution is regular. 
Writing $p = \alpha + \mathrm{i} \, \beta$, the last condition means 
$|\alpha| > |\beta|$. 
This solution appeared in \cite{Gils+Nimm90} (with the opposite inequality  
$|\alpha| < |\beta|$, since our $p$ corresponds to $\mathrm{i} \, p$ 
in that work). 
Fig.~\ref{fig:BKP1lump} shows a plot. 
The factor $4 \, |p|^2$ in (\ref{BKP1lump}) drops out 
in the passage to $\phi$ and can thus be omitted. 
Setting $t_{2n+1}=0$ for $n>2$, the maximum value of $u = \phi_{t_1}$ for the above solution 
is given by $u_{\mathrm{max}} = 4 \, \mathrm{Im}(p^2)^2/\mathrm{Re}(p^2)$ 
and the maximum moves, in `time' $t_5$, according to
\be
   t_1 = 5 \, |p|^4 \, t_5 + \mathrm{Re}(c) 
         - \frac{\mathrm{Re}(p^2)}{\mathrm{Im}(p^2)} \, \mathrm{Im}(c) \, , 
         \qquad
   t_3 = \frac{10}{3} \, \mathrm{Re}(p^2) \, t_5 
         - \frac{\mathrm{Im}(c)}{3 \, \mathrm{Im}(p^2)} \; .
\ee
The solution has two minima with 
$u_{\mathrm{min}} = - \mathrm{Im}(p^2)/(2 \mathrm{Re}(p^2))$, located 
symmetrically with respect to the maximum. See also Fig.~\ref{fig:BKP1lump}. 
\hfill $\square$
\end{example}

\begin{figure}[t] 
\begin{center} 
\resizebox{12cm}{!}{
\includegraphics{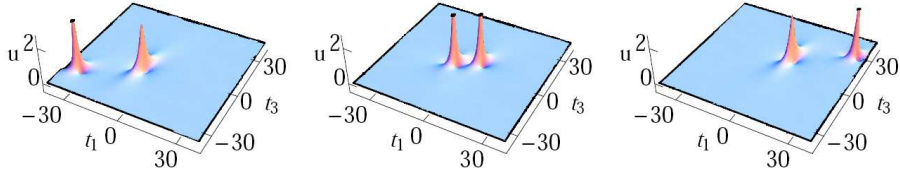}
}
\end{center} 
\caption{A 2-lump solution to the BKP equation. Plot of $u = \phi_{t_1}$ 
at $t_5=-50,0,50$ (and $t_7,t_9,\ldots =0$) for the solution in example~\ref{ex:BKPsuperposition} 
with $C = 0$ and $p_1 = \frac{1}{2} + \frac{\mathrm{i}}{3}$, 
$p_2 = \frac{1}{3} + \frac{\mathrm{i}}{4}$. The two lumps never merge but seem 
to exchange their identities at a certain minimal separation. 
   \label{fig:2nonintBKPlumps} }
\end{figure}

\begin{example}
\label{ex:BKPsuperposition}
Let 
\be
     L_i = \left( \begin{array}{cccc} 0 & p_i & 0 & 0 \\ 
                               -p_i & 0 & 0 & 0 \\
                                0 & 0 & 0 & p_i^\ast \\
                                0 & 0 & -p_i^\ast & 0  
            \end{array} \right) \, , \qquad 
     V_i = \left( \begin{array}{c} 1 \\ 1 \\ 1 \\ 1
                      \end{array} \right) \qquad   i=1,2 \; .
\ee
In the preceding example we have seen that these data 
determine single BKP lumps, and the corresponding $K_i$ are obtained from (\ref{case2_N=2_K}) 
and (\ref{case2_N=4_K12}). 
The superposition condition (\ref{case2_super}) is then solved by
\be
   K_{12} = \left( \begin{array}{cccc} 
            - \frac{1}{p_1+p_2} &  \frac{1}{p_2-p_1} &
            - \frac{1}{p_1+p_2^\ast} & \frac{1}{p_2^\ast-p_1} \\
            \frac{1}{p_1-p_2} & \frac{1}{p_1+p_2} & \frac{1}{p_1-p_2^\ast}
            & \frac{1}{p_1+p_2^\ast} \\
            - \frac{1}{p_1^\ast+p_2} & \frac{1}{p_2-p_1^\ast} &
            - \frac{1}{p_1^\ast+p_2^\ast} & \frac{1}{p_2^\ast-p_1^\ast} \\
            \frac{1}{p_1^\ast-p_2} & \frac{1}{p_1^\ast+p_2} & 
            \frac{1}{p_1^\ast-p_2^\ast} & \frac{1}{p_1^\ast+p_2^\ast}
            \end{array} \right)  \; .
\ee
All this determines BKP 2-lump solutions via (\ref{case2_BKP_tau}) and 
Fig.~\ref{fig:2nonintBKPlumps} displays an example. 
\hfill $\square$
\end{example}

\begin{example}
\label{ex:CKP_L3x3}
Let $N=3$. The general skew-symmetric $3 \times 3$ matrix is 
\be
  L = \left( \begin{array}{ccc} 0 & p_1 & p_3 \\ 
                             -p_1 &  0  & p_2 \\
                             -p_3 & -p_2& 0 
       \end{array} \right)  
\ee
with constants $p_1,p_2,p_3$. Without loss of generality we may set $V^\intercal = (1,1,1)$. 
From $\tr(L^2 (Q-I_3))=0$ we obtain the constraint $p_1 p_2 - p_1 p_3 - p_2 p_3 = 0$, 
which we solve for $p_3 = p_1p_2/(p_1+p_2)$, assuming $p_1+p_2 \neq 0$. 
The solution to (\ref{I+[L,K]=VV^t}) is then given by 
\be	
 K = k_1 \, I_3 
   + \left( \begin{array}{ccc} 
     k_2 \, (1 + \frac{p_2}{p_1}) ( \frac{p_1}{p_2}-\frac{p_2}{p_1}) 
     - \frac{1}{p_1} - \frac{1}{p_2}  
    & k_2 \, \frac{p_2}{p_1} & - k_2 \, (1+\frac{p_2}{p_1}) \\ 
    k_2 \, \frac{p_2}{p_1} 
    & k_2 \, ( \frac{p_1}{p_2}+\frac{p_1}{p_1+p_2} ) - \frac{1}{p_2}
    & k_2 \\
    - k_2 \, (1+\frac{p_2}{p_1}) & k_2 & 0
            \end{array} \right) 
\ee
with arbitrary constants $k_1,k_2$. In the CKP case, the resulting function $\tau$ 
cannot be real and regular (since e.g. at $t_3 = t_5 = \ldots = 0$ it is a third order 
polynomial in $t_1$). 
In the BKP case, it is not really justified to use (\ref{case2_BKP_tau_pre}), since it has 
been derived under the condition $\det(L) \neq 0$, but here $N$ is odd 
and thus $\det(L) = 0$ (because $L$ is skew-symmetric). Nevertheless, (\ref{case2_BKP_tau_pre}) 
yields a solution, though an uninteresting one, since 
$\tau = - \mathfrak{p}^2$ with $\mathfrak{p}$ linear in $t_1,t_3,\ldots$. 
We should rather go back to (\ref{case2_BKP_C^t=-C}) and (\ref{comm_L}), but it turns out 
that these equations cannot both be satisfied non-trivially in the case under consideration. 
\hfill $\square$
\end{example}

\section{Conclusions}
\label{sec:conclusions}
\setcounter{equation}{0}
The odd KP system studied in this work is a system of two PDEs for two dependent variables, 
$\phi$ and $\theta$, taking values in any associative (and typically noncommutative) algebra $\cA$. 
We have shown how this is embedded in the KP hierarchy, if the latter is expressed with the 
help of an auxiliary dependent variable (related to $\theta$).
In particular, this allowed to adapt a construction of exact solutions for the KP hierarchy 
to the odd KP system (and the corresponding hierarchy). We further demonstrated how this 
can be exploited to generate solutions to the BKP and the CKP equation (and their hierarchies). 
In the latter cases we worked out only comparatively simple examples of solutions explicitly. 
The general formulae, however, involve constant matrices of arbitrary size, with little restrictions, 
and with certain choices they may lead to further interesting solutions. 

If $\cA$ is commutative, the odd KP system admits reductions to the BKP and the CKP equation. 
In the noncommutative case, these reductions lead to severely constrained extensions of 
these equations. Nevertheless, they turned out to be helpful since they allowed to uncover 
some properties of the commutative equations (see the relations with the KdV hierarchy in 
sections~\ref{subsec:BKP&ncKdV} and \ref{subsec:CKP&ncKdV}) 
that are hardly recognizable without the step into the noncommutative realm. 
Whereas the CKP equation possesses a natural noncommutative generalization, though as a system 
with two dependent variables, nothing comparable has been found for BKP. 
We also considered some other reductions of the odd KP system with noncommutative $\cA$ and 
obtained in particular a noncommutative version of a coupled system of Kaup-Kupershmidt and 
Sawada-Kotera type. The odd KP system, (\ref{oddKP1}) and (\ref{oddKP2}) with noncommutative 
$\cA$, and its reductions, have not been studied previously according to our knowledge. 

Furthermore, we presented different formulations of the odd KP hierarchy (with noncommutative $\cA$), 
and derived in particular a functional representation of a linear system for the whole hierarchy. 
We verified that all these hierarchy formulations possess the odd KP system as their simplest member. 
Because of the KP hierarchy origin and the hierarchy property one then expects the 
equivalence of all these hierarchy formulations, but a formal proof would nicely 
complement this work. 

The relation between KP and BKP (CKP) via odd KP shows that a subhierarchy can admit a symmetry 
reduction that does not extend to a symmetry reduction of the whole hierarchy. This suggests 
to take a corresponding look at other subhierarchies of KP, and moreover subhierarchies 
of other hierarchies. Besides the odd KP there is evidently also an ``even KP'' subhierarchy 
of the KP hierarchy. In the GDS formulation, this means restricting (\ref{KP-GD_Lax}) to 
even-numbered variables. We shall report on this elsewhere.

\renewcommand{\theequation} {\Alph{section}.\arabic{equation}}
\renewcommand{\thesection} {\Alph{section}}
\newtheorem{lemmaA}{Lemma}[section]

\section*{Appendix A: Proof of Theorem~\ref{theorem:oddKP_funct_linsys} }
\setcounter{section}{1}
\setcounter{equation}{0}
For the evaluation of the bilinear identity (\ref{bilinear-id}), we will use the 
residue formula (which is Lemma~6.3.2 in \cite{Dick03})
\be
    \res \: \frac{f(z)}{1-\lambda z} = \lambda^{-1} \, f_{<0}(\lambda^{-1}) \,  ,
             \label{res_lemma}
\ee
where $f_{<0}(z) = \sum_{n=1}^{+\infty} f_n \, z^{-n}$. In the following, a prime denotes a 
partial derivative with respect to $t_1$, hence e.g. $\phi' := \phi_{t_1}$.
\vskip.2cm

\begin{lemmaA} 
The following are consequences of the bilinear identity (\ref{bilinear-id}). 
We have 
\be
   \tw_2 = \tilde{\theta} + \frac{1}{2} (\phi' + \phi^2) \, ,   \label{tw2-theta}
\ee 
and
\be
    w_{2[\lambda]}(\lambda^{-1}) \, \tw(\lambda^{-1}) = F(\lambda) \, ,  \label{wtw-F}
\ee
with $F(\lambda)$ defined in (\ref{F(la)}). Furthermore, 
\be
  &&\hspace{-1.2cm}  \left( w'(\lambda^{-1}) + \lambda^{-1} w(\lambda^{-1}) \right)_{2[\lambda]} \, \tw(\lambda^{-1})
  = \lambda^{-1} F(\lambda)^2 - \frac{\lambda}{2} (\check{\theta}_{2[\lambda]} - \check{\theta}) 
    + \frac{\lambda}{4} [\phi,\phi_{2[\lambda]}] \, , 
                        \label{wtw1}   \\
  &&\hspace{-1.2cm}  \mu^{-1} \, w_{2[\lambda]+2[\mu]}(\mu^{-1})  \, \tw(\mu^{-1})
    - \lambda^{-1} \, w_{2[\lambda]+2[\mu]}(\lambda^{-1}) \, \tw(\lambda^{-1})
    = (\mu^{-1}-\lambda^{-1}) \, F(\lambda,\mu) \, ,
                        \label{wtw2}  \\
  &&\hspace{-1.2cm} \mu^{-1} \left( w'(\mu^{-1}) 
  + \mu^{-1} w(\mu^{-1}) \right)_{2[\lambda]+2[\mu]} \, \tw(\mu^{-1})
  - \lambda^{-1} \left( w'(\lambda^{-1})+\lambda^{-1}w(\lambda^{-1}) \right)_{2[\lambda]+2[\mu]} \, \tw(\lambda^{-1})  
                          \nonumber \\
  && = (\mu^{-2}-\lambda^{-2}) \, F(\lambda,\mu)^2
    - \frac{1}{2} \frac{\lambda-\mu}{\lambda+\mu} (\check{\theta}_{2[\lambda]+2[\mu]}
    -\check{\theta}) + \frac{1}{4} \frac{\lambda-\mu}{\lambda+\mu} \, [\phi,\phi_{2[\lambda]+2[\mu]}] \, , 
                        \label{wtw3}
\ee
where $\check{\theta} := \tilde{\theta} + \frac{1}{2} \, \phi'$, and
\be
    F(\lambda,\mu) := 1 - \frac{1}{2} \frac{\lambda \mu}{\lambda+\mu} \, (\phi_{2[\lambda]+2[\mu]} - \phi) 
                = \frac{1}{\lambda+\mu} \, \left( \mu \, F_{2[\mu]}(\lambda) + \lambda \, F(\mu) \right) \; .
               \label{Flamu}
\ee
\end{lemmaA}
\noindent
\textit{Proof:} Taking the derivative of (\ref{bilinear-id2}) with respect to $s_1$ and then
setting $\tbs = \tbt$, leads to
\bez
    0 = \res \left( w'(z) \, \tw(z) + z \, w(z) \, \tw(z) \right) = w_1' + w_2 + w_1 \tw_1 + \tw_2 \; .
\eez
Using (\ref{tw1=-w1}) and (\ref{w2-theta}), this becomes (\ref{tw2-theta}). 
With the help of the identities
\bez
    \exp\left(\pm \sum_{n \geq 1} \frac{(\lambda z)^n}{n} \right) = (1-\lambda z)^{\mp 1} \, , \quad
    \mathrm{hence}  \qquad
    \exp\left(2 \sum_{n \geq 1} \frac{(\lambda z)^{2n-1}}{2n-1} \right) 
       = \frac{1 + \lambda z}{1-\lambda z}  \, ,
\eez
(\ref{bilinear-id2}) for $\tbs = \tbt +2[\lambda]$ becomes
\bez
    0 = \res \left( \frac{1+\lambda z}{1-\lambda z} \, w_{2[\lambda]}(z) \tw(z) \right)
      = 2 \lambda^{-1} w_{2[\lambda]}(\lambda^{-1}) \, \tw(\lambda^{-1}) - 2 \lambda^{-1} 
        - (w_1)_{2[\lambda]} - \tw_1 \, ,
\eez 
which is (\ref{wtw-F}). Next we differentiate (\ref{bilinear-id2}) with respect to $s_1$ and then set 
$\tbs = \tbt+2[\lambda]$ to obtain 
\bez
    \res \left( \frac{1+\lambda z}{1-\lambda z} \,
    \left( w'_{2[\lambda]}(z) \, \tw(z) + z \, w_{2[\lambda]}(z) \, \tw(z) \right) \right) = 0 \; .
\eez
Elaborated with the help of (\ref{res_lemma}), and using (\ref{tw1=-w1}), (\ref{w2-theta}) 
and (\ref{tw2-theta}), this results in (\ref{wtw1}).
Furthermore, setting $\tbs=\tbt+2[\lambda]+2[\mu]$ in (\ref{bilinear-id2}), we obtain
\bez
    \res \left( \frac{1+z \lambda}{1-z \lambda} \, \frac{1+z \mu}{1-z\mu} \, w_{2[\lambda]+2[\mu]}(z) \, \tw(z) \right) = 0 \; .
\eez
With the partial fraction decomposition 
\bez
   \frac{1+\lambda z}{1-\lambda z} \, \frac{1+\mu z}{1-\mu z} = 1 + 2 \frac{\lambda+\mu}{\lambda-\mu}
        \left( \frac{1}{1-\lambda z} - \frac{1}{1-\mu z} \right) \, ,
\eez
this results in (\ref{wtw2}). Finally, we differentiate (\ref{bilinear-id2}) with respect to $s_1$,  
and then set $\tbs=\tbt+2[\lambda]+2[\mu]$ to obtain
\bez
    \res \left( \frac{1+z \lambda}{1-z\lambda} \, \frac{1+z \mu}{1-z \mu} \left( w'_{2[\lambda]+2[\mu]}(z) \, \tw(z)
    + z \, w_{2[\lambda]+2[\mu]}(z) \, \tw(z) \right) \right) = 0 \, , 
\eez
which evaluates to (\ref{wtw3}).
\hfill $\square$
\vskip.2cm

\noindent
\textbf{Proof of the theorem:} 
With the help of (\ref{wtw-F}), we can write (\ref{wtw1}) in the form
\bez
    \left( w'(\lambda^{-1}) + \lambda^{-1} w(\lambda^{-1}) \right)_{2[\lambda]}
  = \left( \lambda^{-1} F(\lambda) 
    - \frac{\lambda}{2} \Big( \check{\theta}_{2[\lambda]} - \check{\theta} 
    - \frac{1}{2} \, [\phi,\phi_{2[\lambda]}] \Big) \, F(\lambda)^{-1} \right) w_{2[\lambda]}(\lambda^{-1}) \; .
\eez
Now we apply a Miwa shift with $2[\mu]$ and then multiply by $\tw(\lambda^{-1})$ from the right to obtain
\bez
    \lefteqn{ \left( w'(\lambda^{-1}) + \lambda^{-1} w(\lambda^{-1}) \right)_{2[\lambda]+2[\mu]} \, \tw(\lambda^{-1})=}
    \hspace*{0cm}&&  \nonumber  \\
    && \Big(\lambda^{-1} F(\lambda) - \frac{\lambda}{2} \Big( \check{\theta}_{2[\lambda]} 
     - \check{\theta} - \frac{1}{2} \, [\phi,\phi_{2[\lambda]}] \Big) \, F(\lambda)^{-1}
    \Big)_{2[\mu]} \, w_{2[\lambda]+2[\mu]}(\lambda^{-1}) \, \tw(\lambda^{-1}) \, .
\eez
Inserting this in (\ref{wtw3}) leads to
\bez
    && \mu^{-1} \left( w'(\mu^{-1}) + \mu^{-1} w(\mu^{-1}) \right)_{2[\lambda]+2[\mu]} 
         \, \tw(\mu^{-1}) \nonumber\\
    && -\Big( \lambda^{-1} F(\lambda)^2 - \frac{\lambda}{2} ( \check{\theta}_{2[\lambda]} - \check{\theta} )
    + \frac{\lambda}{4} \, [\phi , \phi_{2[\lambda]}] \Big)_{2[\mu]} \, F_{2[\mu]}(\lambda)^{-1} \lambda^{-1}
    w_{2[\lambda]+2[\mu]}(\lambda^{-1}) \, \tw(\lambda^{-1}) \nonumber\\
    &&= \frac{\lambda^2-\mu^2}{\lambda^2 \mu^2} \, F(\lambda,\mu)^2 - \frac{1}{2} \frac{\lambda-\mu}{\lambda+\mu} \, 
    (\check{\theta}_{2[\lambda]+2[\mu]} - \check{\theta}) + \frac{1}{4} \frac{\lambda-\mu}{\lambda+\mu} \, 
    [\phi,\phi_{2[\lambda]+2[\mu]}] \; .
\eez
Next we use (\ref{wtw2}) to eliminate the factor $\lambda^{-1} w_{2[\lambda]+2[\mu]}(\lambda^{-1}) \, \tw(\lambda^{-1})$, 
\bez
    &&  w'_{2[\lambda]+2[\mu]}(\mu^{-1}) \, \tw(\mu^{-1}) 
       + \Big( \mu^{-1} F(\lambda) - \lambda^{-1} F(\lambda)^2 
       + \frac{\lambda}{2} ( \check{\theta}_{2[\lambda]} - \check{\theta} )  \nonumber \\
    && - \frac{\lambda}{4} \, [\phi,\phi_{2[\lambda]}] \Big)_{2[\mu]} \, 
         F_{2[\mu]}(\lambda)^{-1} \, w_{2[\lambda]+2[\mu]}(\mu^{-1}) \, \tw(\mu^{-1})
                                                                              \nonumber\\
   &=& \frac{\lambda-\mu}{2} \, \Big( (\check{\theta}_{2[\lambda]} 
        - \check{\theta})_{2[\mu]} \, F_{2[\mu]}(\lambda)^{-1} \, F(\lambda,\mu)
      - \frac{\mu}{\lambda+\mu} \, (\check{\theta}_{2[\lambda]+2[\mu]} - \check{\theta}) \Big)
                                                                               \nonumber\\
    &&  + \frac{\mu}{4} \, \frac{\lambda-\mu}{\lambda+\mu} \, [\phi,\phi_{2[\lambda]+2[\mu]}] 
        - \frac{\lambda-\mu}{4} \, [\phi_{2[\mu]},\phi_{2[\lambda]+2[\mu]}] \, F_{2[\mu]}(\lambda)^{-1} \, F(\lambda,\mu)
                                                                              \nonumber \\
    && + \frac{\lambda-\mu}{\lambda^2 \mu} \left( (\lambda+\mu) F(\lambda,\mu) - \mu \, F_{2[\mu]}(\lambda) \right) \, F(\lambda,\mu)
         \nonumber \\
   &=& \frac{\lambda-\mu}{\lambda \mu} \, F(\mu) \, F(\lambda,\mu)
    + \frac{1}{2} \, \frac{\lambda-\mu}{\lambda+\mu} \Big(
      \lambda \, ( \check{\theta}_{2[\lambda]} - \check{\theta} )_{2[\mu]} \, F_{2[\mu]}(\lambda)^{-1} \, F(\mu)
    - \mu \, (\check{\theta}_{2[\mu]}-\check{\theta})                    \nonumber\\
    && + \frac{\mu}{2} \, [\phi,\phi_{2[\lambda]+2[\mu]}] - \frac{\lambda+\mu}{2} \, 
    [\phi_{2[\mu]},\phi_{2[\lambda]+2[\mu]}] \, F_{2[\mu]}(\lambda)^{-1} \, F(\lambda,\mu) \Big)
           \nonumber \\
   &=& \frac{\lambda-\mu}{\lambda + \mu} \, \Big( 
     F(\mu) \, \left( (\lambda^{-1} + \mu^{-1}) \, F(\lambda,\mu) - \mu^{-1} F(\mu) \right) 
    - \frac{\mu}{4} \, [\phi_{2[\mu]} - \phi , \phi_{2[\lambda]+2[\mu]} - \phi_{2[\mu]}] \, \nonumber\\
    && + \frac{\lambda}{2} ( \check{\theta}_{2[\lambda]} - \check{\theta} )_{2[\mu]} \, F_{2[\mu]}(\lambda)^{-1} \,  F(\mu)
    - \frac{\lambda}{4} \, [\phi_{2[\mu]} , \phi_{2[\lambda]+2[\mu]}] \, F_{2[\mu]}(\lambda)^{-1} \, F(\mu)
                                            \nonumber\\
    && + [w'_{2[\mu]}(\mu^{-1}) + \mu^{-1} w_{2[\mu]}(\mu^{-1})] \, \tw(\mu^{-1})   \Big) 
                                            \nonumber\\
   &=& \frac{\lambda-\mu}{\lambda + \mu} \, \Big( \frac{1}{\lambda} F(\lambda) 
       + \frac{\lambda}{2} ( \check{\theta}_{2[\lambda]} - \check{\theta} ) \, F(\lambda)^{-1} 
       - \frac{\lambda}{4} \, [\phi , \phi_{2[\lambda]}] \, F(\lambda)^{-1} \Big)_{2[\mu]} \, F(\mu)
                                            \nonumber\\
    && + \frac{\lambda-\mu}{\lambda + \mu} \, \left( w'(\mu^{-1}) 
       + \mu^{-1} w(\mu^{-1}) \right)_{2[\mu]} \, \tw(\mu^{-1})        \, ,
\eez
taking account of (\ref{Flamu}), (\ref{wtw1}), 
$(\lambda^{-1} + \mu^{-1}) \, F(\lambda,\mu) - \mu^{-1} F(\mu) = \lambda^{-1} F_{2[\mu]}(\lambda)$, and 
\bez
[F(\mu) , F_{2[\mu]}(\lambda)] = \frac{\lambda \mu}{4} \, [\phi_{2[\mu]} - \phi , \phi_{2[\lambda]+2[\mu]} - \phi_{2[\mu]}]
   \; .
\eez
Now we use (\ref{wtw-F}) to replace the factor $F(\mu)$, divide by $\tw(\mu^{-1})$, and then apply 
a Miwa shift with $-2[\mu]$ to obtain
\bez
    && \frac{\lambda+\mu}{\lambda-\mu} \Big[ w'_{2[\lambda]}(\mu^{-1}) 
       + \Big( \mu^{-1} F(\lambda) - \lambda^{-1} F(\lambda)^2
       + \frac{\lambda}{2}(\check{\theta}_{2[\lambda]} - \check{\theta}) 
     - \frac{\lambda}{4} \, [\phi,\phi_{2[\lambda]}] \Big) \, F(\lambda)^{-1} 
         \, w_{2[\lambda]}(\mu^{-1}) \Big]                \nonumber \\
    &=& w'(\mu^{-1}) + \mu^{-1} w(\mu^{-1})  
       + \Big( \frac{1}{\lambda} \, F(\lambda)
       + \frac{\lambda}{2} ( \check{\theta}_{2[\lambda]} - \check{\theta} ) \, F(\lambda)^{-1}
       - \frac{\lambda}{4} \, [\phi , \phi_{2[\lambda]}] \, F(\lambda)^{-1} \Big) \, 
          w(\mu^{-1}) \; .
\eez
Setting $\mu=z^{-1}$, after some rearrangements this takes the form
\bez
 && \frac{1+\lambda z}{1-\lambda z} \, \left( w'_{2[\lambda]}(z) + z \, w_{2[\lambda]}(z) \right) + w'(z) + z \, w(z) 
     - \frac{1}{\lambda} \, F(\lambda)
     \Big( \frac{1+\lambda z}{1-\lambda z} \, w_{2[\lambda]}(z) -w(z) \Big) \nonumber \\
 && + \frac{\lambda}{2} \Big( \check{\theta}_{2[\lambda]} - \check{\theta} 
       - \frac{1}{2} \, [\phi,\phi_{2[\lambda]}] \Big) \, F(\lambda)^{-1} \Big(
         \frac{1+\lambda z}{1-\lambda z} \, w_{2[\lambda]}(z) + w(z) \Big) = 0 \; .
\eez
Multiplying by $e^{\txi(\tbt,z)}$ and using 
$\psi_{2[\lambda]} = w_{2[\lambda]}(z) \, \frac{1+\lambda z}{1-\lambda z} \, e^{\txi(\tbt,z)}$, 
we arrive at (\ref{fun_oddKP}).
\hfill $\square$

\section*{Appendix B: A determinant identity}
\setcounter{section}{2}
\setcounter{equation}{0}
According to (2.90) in \cite{Hiro04}, we have
\be
   \det \left( \begin{array}{cc} z & V^\intercal \\
                              -V & A \end{array} \right) 
 = \det(A) \, z + \sum_{i,j=1}^N \Delta_{i,j} \, v_i \, v_j \, ,
\ee
where $A$ is an $N \times N$ matrix, $\Delta_{i,j}$ is the cofactor with 
respect to the component $A_{i,j}$ of $A$, $z$ a parameter, and $v_i$, $i=1,\ldots,N$, 
are the components of a vector $V$. If $N$ is odd and $A$ skew-symmetric, 
then $\det(A)=0$ and thus
\be
   \det \left( \begin{array}{cc} z & V^\intercal \\
                              -V & A \end{array} \right) 
 = \sum_{i,j=1}^N \Delta_{i,j} \, v_i \, v_j \, ,
\ee
which is thus independent of $z$. Since 
\be
   \det \left( \begin{array}{cc} 1 & V^\intercal \\
                              -V & A \end{array} \right) 
 =  \det \left( \begin{array}{cc} 1 & V^\intercal \\
                              0 & A + V V^\intercal \end{array} \right) 
 = \det( A + V V^\intercal ) \, ,
\ee
we obtain
\be
   \det( A + V V^\intercal ) = \det \left( \begin{array}{cc} 0 & V^\intercal \\
                              -V & A \end{array} \right) \, ,
\ee
which is the determinant of a skew-symmetric matrix, and thus the square of the Pfaffian 
of this matrix.

\section*{Appendix C: Reality conditions}
\setcounter{section}{3}
\setcounter{equation}{0}
In order to obtain \emph{real} solutions to the BKP or CKP hierarchy from the matrix linear 
system in section~\ref{sec:oddKPsol} with \emph{complex} matrices, a reality condition is 
needed. 

\begin{proposition}
Let $T$ be a constant invertible $N \times N$ matrix with the properties 
\be
   T^\ast = T^\intercal = T^{-1}   \label{T-cond}
\ee
(where $T^\ast$ denotes the complex conjugate of $T$). Let 
$C,K,L$ be constant complex $N \times N$ matrices and $V$ an $N$-vector satisfying 
\be
   C^\ast = T C T^{-1} \, , \quad
   K^\ast = T K T^{-1} \, , \quad
   L^\ast = T L T^{-1} \, , \quad
   V^\ast = T V  \; .                 \label{CKLV_real-cond}
\ee
The function $\tau$ given by (\ref{case1_CKP_sol}),(\ref{case1_BKP_tau}), (\ref{case2_CKP_tau})
or (\ref{case2_BKP_tau}) in terms of $(C,K,L,V)$ (subject to the corresponding rank one condition 
(\ref{rank(L^tK+KL)=1}) or (\ref{I+[L,K]=VV^t}), and $C^\intercal = C$, respectively $C^\intercal = -C$), 
is then \emph{real}. 
\end{proposition}
\noindent
\textit{Proof:} 
The assertion is easily verified. (\ref{T-cond}) ensures the 
compatibility of (\ref{CKLV_real-cond}) with $C^\intercal = \pm C$, 
(\ref{rank(L^tK+KL)=1}) and (\ref{I+[L,K]=VV^t}). 
\hfill $\square$
\vskip.2cm

If $N=2n$, then 
\be
   T = \left( \begin{array}{cc} 0 & I_n \\ I_n & 0 \end{array} \right) \, ,
                  \label{T-choice}
\ee
where $I_n$ is the $n \times n$ unit matrix, satisfies the conditions (\ref{T-cond}). 
Decomposing the matrix $L$ into $n \times n$ blocks, (\ref{CKLV_real-cond}) leads to
\be
   L = \left( \begin{array}{cc} L_1 & L_{12} \\ L_{12}^\ast & L_1^\ast  \end{array} \right) 
        \; . 
\ee
In section~\ref{sec:oddKPsol} we presented examples with such conjugate diagonal blocks 
(and $L_{12} =0$).


\begin{thebibliography}{10}

\bibitem{March88}
Marchenko V 1988 {\em Nonlinear {E}quations and {O}perator {A}lgebras\/}
  Mathematics and Its Applications (Dordrecht: Reidel)

\bibitem{Dorf+Foka92}
Dorfman I and Fokas A 1992 Hamiltonian theory over noncommutative rings and
  integrability in multidimensions {\em J. Math. Phys.\/} {\bf 33} 2504--2514

\bibitem{Olve+Soko98assoc}
Olver P and Sokolov V 1998 Integrable evolution equations on associative
  algebras {\em Commun. Math. Phys.\/} {\bf 193} 245--268

\bibitem{Kupe00}
Kupershmidt B 2000 {\em {KP} or m{KP}\/} ({\em Mathematical Surveys and
  Monographs\/} vol~78) (Providence: American Math. Soc.)

\bibitem{Carl+Schi99}
Carl B and Schiebold C 1999 Nonlinear equations in soliton physics and operator
  ideals {\em Nonlinearity\/} {\bf 12} 333--364

\bibitem{DMH08bidiff}
Dimakis A and M\"uller-Hoissen F 2008 Bidifferential graded algebras and
  integrable systems {\em arXiv:0805.4553\/}

\bibitem{Sawa+Kote74}
Sawada K and Kotera T 1974 A method for finding {N}-soliton solutions of the
  {K.d.V.} equation and {K.d.V.}-like equation {\em Progr. Theor. Phys.\/} {\bf
  51} 1355--1367

\bibitem{DKM81TII}
Date E, Kashiwara M and Miwa T 1981 Vertex operators and $\tau$ functions
  transformation groups for soliton equations, {II} {\em Proc. Japan Acad. Ser.
  A Math. Sci.\/} {\bf 57} 387--392

\bibitem{DJKM82TIV}
Date E, Jimbo M, Kashiwara M and Miwa T 1982 Transformation groups for soliton
  equations. {IV}. {A} new hierarchy of soliton equations of {KP}-type {\em
  Physica\/} {\bf 4D} 343--365

\bibitem{DJKM81TVI}
Date E, Jimbo M, Kashiwara M and Miwa T 1981 {KP} hierarchies of orthogonal and
  symplectic type -- {T}ransformation groups for soliton equations {VI} -- {\em
  J. Phys. Soc. Japan\/} {\bf 50} 3813--3818

\bibitem{DJKM82RIMS}
Date E, Jimbo M, Kashiwara M and Miwa T 1982 Transformation groups for soliton
  equations -- {E}uclidean {L}ie algebras and reduction of the {KP} hierarchy
  -- {\em Publ. RIMS\/} {\bf 18} 1077--1110

\bibitem{DJKM82TV}
Date E, Jimbo M, Kashiwara M and Miwa T 1982 Quasi-periodic solutions of the
  orthogonal {KP} equation -- {T}ransformation groups for soliton equations {V}
  -- {\em Publ. RIMS\/} {\bf 18} 1111--1119

\bibitem{DJM83}
Date E, Jimbo M and Miwa T 1983 Method for generating discrete soliton
  equations. {IV} {\em J. Phys. Soc. Japan\/} {\bf 52} 761--765

\bibitem{DKJM83}
Date E, Kashiwara M, Jimbo M and Miwa T 1983 Transformation groups for soliton
  equations {\em Non-linear {I}ntegrable {S}ystems -- {C}lassical {T}heory and
  {Q}uantum {T}heory\/} ed Jimbo M and Miwa T (Singapore: World Scientific) pp
  39--119

\bibitem{Kono+Dubr84}
Dubrovsky V and Konopelchenko B 1984 Some new integrable nonlinear evolution
  equations in $2 + 1$ dimensions {\em Phys. Lett. A\/} {\bf 102} 15--17

\bibitem{Hiro86}
Hirota R 1986 Reduction of soliton equations in bilinear form {\em Physica D\/}
  {\bf 18} 161--170

\bibitem{Hiro89a}
Hirota R 1989 Soliton solutions to the {BKP} equations. {I}. {T}he {P}faffian
  technique {\em J. Phys. Soc. Japan\/} {\bf 58} 2285--2296

\bibitem{Hiro89b}
Hirota R 1989 Soliton solutions to the {BKP} equations. {II}. {T}he integral
  equation {\em J. Phys. Soc. Japan\/} {\bf 58} 2705--2712

\bibitem{Shiota89}
Shiota T 1989 Prym varieties and soliton equations {\em Infinite-dimensional
  Lie Algebras and Groups\/} ({\em Adv. Ser. Math. Phys.\/} vol~7) ed Kac V
  (Singapore: World Scientific) pp 407--448

\bibitem{Nimm90}
Nimmo J 1990 Hall-{L}ittlewood symmetric functions and the {BKP} equation {\em
  J. Phys. A: Math. Gen.\/} {\bf 23} 751--760

\bibitem{Gils+Nimm90}
Gilson C and Nimmo J 1990 Lump solutions of the {BKP} equation {\em Phys. Lett.
  A\/} {\bf 147} 472--476

\bibitem{Hiro+Ohta91}
Hirota R and Ohta Y 1991 Hierarchies of coupled soliton equations. {I} {\em J.
  Phys. Soc. Japan\/} {\bf 60} 798--809

\bibitem{You91}
You Y 1991 {DKP} and {MDKP} hierarchy of soliton equations {\em Physica D\/}
  {\bf 50} 429--462

\bibitem{Taim91}
Taimanov I 1991 Prym's theta-function and hierarchies of nonlinear equations
  {\em Math. Notes\/} {\bf 50} 723--730

\bibitem{Chow+Dasg91a}
Chowdhury A and Das~Gupta N 1991 Commutative differential operator and a class
  of solutions for the {BKP} equation on the singular rational curve {\em J.
  Math. Phys.\/} {\bf 32} 3473--3475

\bibitem{Chow+Dasg91b}
Chowdhury A and Dasgupta N 1991 Pseudo-differential operators, {BKP} equation
  and {W}eierstrass {$P(x)$} function {\em J. Phys. A: Math. Gen.\/} {\bf 24}
  L1145--L1148

\bibitem{Jarv+Yung93}
Jarvis P and Yung C 1993 Symmetric functions and the {KP} and {BKP} hierarchies
  {\em J. Phys. A: Math. Gen.\/} {\bf 26} 5905--5922

\bibitem{BJY93}
Baker T, Jarvis P and Yung C 1993 Hirota polynomials for the {KP} and {BKP}
  hierarchies {\em Lett. Math. Phys.\/} {\bf 29} 55--62

\bibitem{Naka+Yama94}
Nakajima T and Yamada H 1994 Basic representations of {$A^{(2)}_{2l}$} and
  {$D^{(2)}_{l+1}$} and the polynomial solutions to the reduced {BKP}
  hierarchies {\em J. Phys. A: Math. Gen.\/} {\bf 27} L171--L176

\bibitem{vandL95}
van~de Leur J 1995 The {A}dler-{S}hiota-van {M}oerbeke formula for the {BKP}
  hierarchy {\em J. Math. Phys.\/} {\bf 36} 4940--4951

\bibitem{Nimm95}
Nimmo J 1995 Darboux transformations from reductions of the {KP} hierarchy {\em
  Nonlinear Evolution Equations and Dynamical Systems. NEEDS '94\/} ed
  Makhankov V, Bishop A and Holm D (Singapore: World Scientific) pp 168--178

\bibitem{Plant+Salam95}
Plant A and Salam M 1995 Schur's {$Q$}-functions and the {BKP} hierarchy {\em
  J. Phys. A: Math. Gen.\/} {\bf 28} L59--L62

\bibitem{Tsuj+Hiro96}
Tsujimoto S and Hirota R 1996 Pfaffian representation of solutions to the
  discrete {BKP} hierarchy in bilinear form {\em J. Phys. Soc. Japan\/} {\bf
  65} 2797--2806

\bibitem{Kac+vanderLeur98}
Kac V and van~der Leur J 1998 The geometry of spinors and the multicomponent
  {BKP} and {DKP} hierarchies {\em The Bispectral Problem\/} ({\em CRM Proc.
  Lect. Notes\/} vol~14) ed Harnad J and Kasman A pp 159--202

\bibitem{vandL01}
van~de Leur J 2001 Matrix integrals and the geometry of spinors {\em J. Nonl.
  Math. Phys.\/} {\bf 8} 288--310

\bibitem{Mana+Mart98}
Ma\~{n}as M and Martinez~Alonso L 1998 From {R}amond fermions to {L}am{\'e}
  equations for orthogonal curvilinear coordinates {\em Phys. Lett. B\/} {\bf
  436} 316--322

\bibitem{Plaz98}
Plaza~Martin F 1998 Prym varieties and the infinite {G}rassmannian {\em Int. J.
  Math.\/} {\bf 9} 75--93

\bibitem{Liu+Mana98}
Liu Q and Ma{\~n}as M 1998 Pfaffian form of the {G}rammian determinant
  solutions of the {BKP} hierarchy {\em solv-int/9806004\/}

\bibitem{DMM99}
Doliwa A, Ma{\~n}as M and Martinez~Alonso L 1999 Generating quadrilateral and
  circular lattices in {KP} theory {\em Phys. Lett. A\/} {\bf 262} 330--343

\bibitem{Lori+Willo99}
Loris I and Willox R 1999 Symmetry reductions of the {BKP} hierarchy {\em J.
  Math. Phys.\/} {\bf 40} 1420--1431

\bibitem{Lori01}
Loris I 2001 Dimensional reductions of {BKP} and {CKP} hierarchies {\em J.
  Phys. A: Math. Gen.\/} {\bf 34} 3447--3459

\bibitem{Gils02}
Gilson C 2002 Generalizing the {KP} hierarchies: {P}faffian hierarchies {\em
  Theor. Math. Phys.\/} {\bf 133} 1663--1674

\bibitem{DFL02}
Dubrovsky V, Formusatik I and Lisitsyn Y 2002 New exact solutions of some
  two-dimensional integrable nonlinear equations via
  $\overline{\partial}$-dressing method {\em Proc. Inst. Math. NAS Ukraine\/}
  {\bf 43} 302--313

\bibitem{Dubr+Lisi02}
Dubrovsky V and Lisitsyn Y 2002 The construction of exact solutions of
  two-dimensional integrable generalizations of {K}aup-{K}uperschmidt and
  {S}awada-{K}otera equations via $\overline{\partial}$-dressing method {\em
  Phys. Lett. A\/} {\bf 295} 198--207

\bibitem{Muse+Verh03}
Musette M and Verhoeven C 2003 {CKP} and {BKP} equations related to the
  generalized quartic {H}{\'e}non-{H}eiles {H}amiltonian {\em Theor. Math.
  Phys.\/} {\bf 137} 1561--1573

\bibitem{Orlov03}
Orlov A 2003 Hypergeometric functions related to {S}chur {$Q$}-polynomials and
  the {BKP} equation {\em Theor. Math. Phys.\/} {\bf 137} 1574--1589

\bibitem{Hiro04}
Hirota R 2004 {\em The {D}irect {M}ethod in {S}oliton {T}heory\/} ({\em
  Cambridge Tracts in Mathematics\/} vol 155) (Cambridge: Cambridge University
  Press)

\bibitem{Nimm+Orlo05}
Nimmo J and Orlov A~Y 2005 A relationship between rational and multi-soliton
  solutions of the {BKP} hierarchy {\em Glasgow Math. J.\/} {\bf 47A} 149--168

\bibitem{Kric06}
Krichever I 2006 A characterization of {P}rym varieties {\em Int. Math. Res.
  Not.\/} {\bf 2006} 81476

\bibitem{HCR06}
He J, Cheng Y and R\"omer R 2006 Solving bi-directional soliton equations in
  the {KP} hierarchy by gauge transformation {\em JHEP\/} {\bf 200603} 103

\bibitem{Hu+Wang06}
Hu X~B and Wang H~Y 2006 Construction of d{KP} and {BKP} equations with
  self-consistent sources {\em Inv. Problems\/} {\bf 22} 1903--1920

\bibitem{Taka06sig}
Takasaki K 2006 Dispersionless {H}irota equations of two-component {BKP}
  hierarchy {\em SIGMA\/} {\bf 2} 057

\bibitem{Taka07}
Takasaki K 2007 Differential {F}ay identities and auxiliary linear problem of
  integrable hierarchies {\em arXiv:0710.5356\/}

\bibitem{Tu07}
Tu M~H 2007 On the {BKP} hierarchy: Additional symmetries, {F}ay identity and
  {A}dler-{S}hiota-van {M}oerbeke formula {\em Lett. Math. Phys.\/} {\bf 81}
  93--105

\bibitem{HWC07}
He J, Wu Z and Cheng Y 2007 Gauge transformations for the constrained {CKP} and
  {BKP} hierarchies {\em J. Math. Phys.\/} {\bf 48} 113519

\bibitem{vandL+Orlov08}
van~de Leur J and Orlov A 2008 Random turn walk on a half line with creation of
  particles at the origin {\em arXiv:0801.0066\/}

\bibitem{Lori99}
Loris I 1999 On reduced {CKP} equations {\em J. Phys. A: Math. Gen.\/} {\bf 15}
  1099--1109

\bibitem{Arat+vandeL05}
Aratyn H and van~de Leur J 2005 The symplectic {K}adomtsev-{P}etviashvili
  hierarchy and rational solutions of {P}ainlev{\'e} {VI} {\em Ann. Inst.
  Fourier\/} {\bf 55}

\bibitem{Matsuno05DP1}
Matsuno Y 2005 Multisoliton solution of the {D}egasperis-{P}rocesi equation and
  their peakon limit {\em Inv. Problems\/} {\bf 21} 1553--1570

\bibitem{HCR07}
He J, Cheng Y and R\"omer R 2007 Two-peak soliton in the {CKP} hierarchy {\em
  Chaos Sol. Fract.\/} {\bf 31}

\bibitem{DMH07Burgers}
Dimakis A and M\"uller-Hoissen F 2007 Burgers and {KP} hierarchies: A
  functional representation approach {\em Theor. Math. Phys.\/} {\bf 152}
  933--947

\bibitem{Gekh+Kasm06}
Gekhtman M and Kasman A 2006 On {KP} generators and the geometry of the {HBDE}
  {\em J. Geom. Phys.\/} {\bf 56} 282--309

\bibitem{DMH06nahier}
Dimakis A and M\"uller-Hoissen F 2006 Nonassociativity and integrable
  hierarchies {\em nlin.SI/0601001\/}

\bibitem{DMH07Ricc}
Dimakis A and M\"uller-Hoissen F 2008 Weakly nonassociative algebras, {R}iccati
  and {KP} hierarchies {\em Generalized Lie Theory in Mathematics, Physics and
  Beyond\/} ed Silvestrov S, Paal E, Abramov V and Stolin A (Berlin: Springer)
  pp 9--27

\bibitem{DMH07Wronski}
Dimakis A and M\"uller-Hoissen F 2007 With a {C}ole-{H}opf transformation to
  solutions of the noncommutative {KP} hierarchy in terms of {W}ronski matrices
  {\em J. Phys. A: Math. Theor.\/} {\bf 40} F321--F329

\bibitem{Dick03}
Dickey L 2003 {\em Soliton {E}quations and {H}amiltonian {S}ystems\/}
  (Singapore: World Scientific)

\bibitem{Kash+Miwa81}
Kashiwara M and Miwa T 1981 The $\tau$ function of the
  {K}adomtsev-{P}etviashvili equation. {T}ransformation groups for soliton
  equations. {I} {\em J. Japan Acad., Ser. A\/} {\bf 57} 342--347

\bibitem{Rama81}
Ramani A 1981 Inverse scattering, ordinary differential equations of
  {P}ainlev{\'e}-type, and {H}irota's bilinear formalism {\em Ann. New York
  Acad. Sci.\/} {\bf 373} 54--–67

\bibitem{Dye+Park01}
Dye J and Parker A 2001 On bidirectional fifth-order nonlinear evolution
  equations, {L}ax pairs, and directionally solitary waves {\em J. Math.
  Phys.\/} {\bf 42} 2567--2589

\bibitem{Dye+Park02}
Dye J and Parker A 2002 A bidirectional {K}aup-{K}upershmidt equation and
  directionally dependent solitons {\em J. Math. Phys.\/} {\bf 43} 4921--4949

\bibitem{Park+Dye02}
Parker A and Dye J 2002 Boussinesq-type equations and ``switching'' solitons
  {\em Proc. Inst. Math. NAS Ukraine\/} {\bf 43} 344--351

\bibitem{Kaup80}
Kaup D 1980 On the inverse scattering problem for cubic eigenvalue problems of
  the class $\psi_{3x} + 6 q \psi_x + 6r \psi = \lambda \psi$ {\em Stud. Appl.
  Math.\/} {\bf 62} 189--216

\bibitem{Verh+Muse03}
Verhoeven C and Musette M 2003 Soliton solutions of two bidirectional
  sixth-order partial differential equations belonging to the {KP} hierarchy
  {\em J. Phys. A: Math. Gen.\/} {\bf 36} L133--L143

\bibitem{DMH04hier}
Dimakis A and M\"uller-Hoissen F 2004 Extension of noncommutative soliton
  hierarchies {\em J. Phys. A: Math. Gen.\/} {\bf 37} 4069--4084

\bibitem{MSS91}
Mikhailov A, Shabat A and Sokolov V 1991 The symmetry approach to
  classification of integrable equations {\em What Is Integrability\/} Springer
  Series in Nonlinear Dynamics ed Zakharov V (Springer) pp 115--184

\bibitem{Four+More01}
Foursov M and Moreno~Maza M 2001 On the relationship between the
  {K}aup-{K}upershmidt and {S}awada-{K}otera equations {\em rapport technique
  LIFL 2001-04, Universit{\'e}� de Lille-I\/}

\bibitem{Four+More02}
Foursov M and Moreno~Maza M 2002 On computer-assisted classification of coupled
  integrable equations {\em J. Symb. Comp.\/} {\bf 33} 647--660

\bibitem{MSY87}
Mikhailov A, Shabat A and Yamilov R 1987 The symmetry approach to the
  classification of nonlinear equations. {C}omplete list of integrable systems
  {\em Russ. Math. Surv.\/} {\bf 42} 1--63

\bibitem{Four99}
Foursov M 1999 {\em On integrable evolution equations in commutative and
  noncommutative variables\/} {PhD} thesis University of Minnesota

\bibitem{Verm02}
Vermaseren J 2002 {\em {FORM} {R}eference {M}anual\/} (Amsterdam: NIKHEF)

\bibitem{Wolf03}
Wolfram S 2003 {\em The {M}athematica Book, 5th ed.\/} (Champain, USA: Wolfram
  Media)

\bibitem{Jimbo+Miwa83}
Jimbo M and Miwa T 1983 Solitons and infinite dimensional {L}ie algebras {\em
  Publ. RIMS\/} {\bf 19} 943--1001

\bibitem{Brus+Ragn83}
Bruschi M and Ragnisco O 1983 On the inversion of the commutation operator {\em
  Lett. Nuovo Cimento\/} {\bf 38} 41--44

\bibitem{Mori98}
Morimoto M 1998 {\em Analytic {F}unctionals on the {S}phere\/} ({\em
  Translations of Mathematical Monographs\/} vol 178) (AMS)

\end{thebibliography}
\end{document}